\documentclass[epj,final]{svjour}
\usepackage{epsfig,pstricks}
\usepackage{graphicx}

\newcommand{\lsim}{\stackrel{\scriptstyle <}{\phantom{}_{\sim}}}
\newcommand{\gsim}{\stackrel{\scriptstyle >}{\phantom{}_{\sim}}}

\begin{document}

\title{Influence of the stiffness of the equation of state and in-medium effects on the cooling of compact stars}
\titlerunning{Stiff EoS and medium effects on cooling of compact stars}

\author{H.~Grigorian\inst{1,2}
\and
D.~N.~Voskresensky\inst{3}
\and
D.~Blaschke\inst{3,4,5}
}
\institute{
Laboratory for Information Technologies,
Joint Institute for Nuclear Research,
Joliot-Curie Str. 6,
141980 Dubna,
Russia
\and
 Department of Physics,
Yerevan State University,
Alek Manukyan Str. 1,
Yerevan 0025,
Armenia
\and
National Research Nuclear University (MEPhI),
Kashirskoe Shosse 31,
115409 Moscow, Russia
\and
Institute for Theoretical Physics,
University of Wroclaw,
Max Born place 9,
50-204 Wroclaw,
Poland
\and
Bogoliubov Laboratory for Theoretical Physics,
Joint Institute for Nuclear Research,
Joliot-Curie Str. 6,
141980 Dubna,
Russia
}
\date{Received: \today / Revised version: date}
%
\abstract{
Measurements of the low masses for the pulsar PSR J0737-3039B, for the companion
of PSR J1756-2251 and for the companion of PSR J0453+1559 on the one hand and 
of the high masses for the pulsars PSR J1614-2230 and PSR J0348-0432 on the other demonstrate the existence of compact stars with masses in a broad range from 1.2 to 2 $M_\odot$. 
The most massive of these objects might be hybrid stars.
To fulfill the  constraint $M_{\rm max}>2~M_{\odot}$ with a reserve we exploit the stiff DD2 hadronic equation of state (EoS) without and with excluded volume (DD2vex) correction, which produce  maximum neutron star masses of $M_{\rm max} = 2.43~M_{\odot}$ and  $2.70~M_{\odot}$, respectively.   
We show that the stiffness of the EoS does not preclude an explanation of the whole set of cooling data within ``nuclear medium cooling" scenario for compact stars by a variation of the star masses. 
We select appropriate proton gap profiles from those exploited in the literature and allow for a variation of the effective pion gap controlling the efficiency of the medium modified Urca process. 
However, we suppress the possibility of pion condensation. 
In general, the stiffer the EoS the steeper a decrease with density of the effective pion gap is required. Results are compared with previously obtained ones for the HDD EoS for which $M_{\rm max} =2.06~M_{\odot}$. 
The  cooling of the compact star in the supernova remnant Cassiopeia A (Cas A) is explained mainly by   an efficient medium modified Urca process. 
To explain a $\gsim 2.5 \%$ decline of the cooling curve for Cas A, as motivated by an analysis  of the ACIS-S instrument data, together with other cooling data exploiting the DD2 EoS a large proton gap at densities $n \lsim 2n_0$ is required vanishing for $n\gsim 2.5~n_0$, where $n_0$ is the
saturation nuclear density.   
A smaller decline, as it follows from an analysis of the HRC-S instrument data, is  explained with many choices of parameters.
With the DD2vex EoS and using an effective pion gap steeper decreasing with the density and/or a proton gap shifted to smaller densities we are also able to reproduce both a strong decline compatible with ACIS-S data and HRC-S instrument data.   
The mass of Cas A is estimated as $1.6 - 1.9~M_{\odot}$, above the value $1.5~M_{\odot}$, which we have  evaluated  with the softer HDD equation of state. 
Different mass choices for the hottest object XMMU J173203.3-344518 are discussed.
We make general remarks also on hybrid star cooling and on its dependence on the stiffness of the hadronic EoS.
}
\maketitle

\section{\label{sec:intro}Introduction}
Experimental data on surface temperatures of compact stars (CS)
provide us with information about the neutrino emissivities of
various processes, depending on the density behaviour of the nucleon-nucleon ($NN$)
interaction amplitude, values and density profiles of the proton
and neutron paring gaps, the heat transport and the equation of
state (EoS) of CS matter.
Recently the situation has been improved with the observation of a
segment of the cooling curve for the young  CS in the remnant of the
historical supernova Cassiopeia A (Cas A) of the year 1680 
\cite{Tananbaum:1999kx,Ashworth:1980vn},
with known age ($335$ yrs by the time of the writing of this manuscript), 
for which the temperature and the rate of cooling have been followed over 
the years since its discovery by the Chandra observatory in the year 2000
\cite{Ho:2009fk,Heinke:2010xy,Page:2010aw,Yakovlev:2010,Shternin:2010qi,Elshamouty:2013nfa}.
The data from ACIS-S and HRC-S instruments  aboard this satellite require the existence of a fast cooling process in the interior of Cas A, with temperature declines of  $\gsim 2.5\%$ and $< 2\%$ per 10 yrs, resp.
On the other hand, the CS cooling model must also
explain the central compact object (CCO) in the supernova remnant  XMMU J173203.3-344518 \cite{Klochkov}, for which the surface temperature has recently
been measured and the mass of the object is estimated.
This object is hotter and older than Cas A, at an age between 10 and 40 kyr.
Moreover, there exists information on surface temperatures of many other CS. 
These data points can be separated in three groups related to slow cooling (see objects 8; 5; 1; 2; 4; A in Fig.~\ref{Fig:Cool1} below), intermediate cooling (3; 6; 7; Cas A; B; E) and rapid cooling (C; D). 
In order to explain the difference in the cooling of the slowly and rapidly cooling objects a three order of magnitude difference in their luminosities is required. 
Therefore, it is not easy to appropriately explain the essentially different surface temperatures of various objects in the hadronic scenario within the so called ``minimal cooling paradigm'', cf. 
 \cite{Page:2010aw,Elshamouty:2013nfa,Page:2009fu,Potekhin:2015qsa}, where the only relevant rapid process is the so called pair-breaking-formation (PBF) process on neutrons paired in the $3P_2$ channel.
The solution of the puzzle might be associated with a strong
medium dependence of cooling inputs, as provided by the density-dependent (and thus neutron star (NS) mass dependent) medium modifications of the $NN$ interaction.
They are caused by the softening of the pion exchange contribution with increasing density for $n\gsim n_0$, where $n_0\simeq 0.16$ fm$^{-3}$ is the nucleon saturation density, and by the density dependent superfluid pairing gaps, see \cite{Migdal:1990vm,Voskresensky:2001fd,Kolomeitsev:2010pm} for details.
The key idea formulated long ago \cite{Voskresensky:1985qg,Voskresensky:1986af} is that the
cooling of various sources should be essentially different due to the difference in their masses.
At that time there prevailed the opinion that all CS masses should be close to
the values for known binary radiopulsars, $1.35 - 1.4~M_{\odot}$.
The recent measurements of the high masses of the pulsars PSR J1614-2230 \cite{Demorest} and
PSR J0348-0432 \cite{Anotoniadis} on the one hand and of the low masses for PSR J0737-3039B \cite{Kramer} for the companion of PSR J1756-2251 \cite{Faulkner,Ferdman:2014rna} 
and the companion of PSR J0453+1559 \cite{Martinez:2015mya}
on the other have provided the proof for the existence of CS masses varying in a broader range, at least from 1.2 to 2 $M_\odot$.

The most efficient processes within the hadronic ``nuclear medium cooling" scenario, cf. \cite{Voskresensky:2001fd}, are the medium modified Urca (MMU) processes, like neutron branch,  
$nn\to npe\bar{\nu}$,  i.e. the modified Urca (MU) processes computed by taking into account pion softening effects \cite{Voskresensky:1986af,Senatorov:1987aa,Migdal:1990vm}, and the
PBF processes $N\to N_{pair}\nu\bar{\nu}$, $N=n$ or $p$ \cite{Flowers:1976ux,Voskresensky:1987hm}.
While being enhanced owing to their one-nucleon nature \cite{Voskresensky:1987hm},
the latter processes are allowed only in the presence of nucleon pairing and should be computed by taking into account the in-medium effects in the weak interaction vertices, cf. \cite{Kolomeitsev:2008mc,Kolomeitsev:2010hr}.
Due to the in-medium effects the vector current contribution proves to be dramatically suppressed in case of $1S_0$ pairing and can be ignored \cite{Leinson:2006gf,Kolomeitsev:2008mc,Kolomeitsev:2010hr}, whereas the axial-vector term is less suppressed \cite{Kolomeitsev:2008mc,Kolomeitsev:2010hr} and should be included both for the PBF processes on neutrons and protons. 
Note that in the minimal cooling scenario one assumes that the emissivity of the PBF process on protons is suppressed by two orders of magnitude compared to that for the PBF process on neutrons since the authors use the free $p\to p\nu\bar{\nu}$ vertices, whereas in matter the  decay may occur through the neutron and neutron hole and the electron and electron hole in the intermediate states of the reaction, cf. \cite{Voskresensky:1987hm,Leinson:2000ks,Voskresensky:2001fd,Kolomeitsev:2008mc,Kolomeitsev:2010hr}.
The pion softening taken into account in MMU processes appears due to the attractive contribution of the nucleon-nucleon hole and delta-nucleon hole diagrams. 
So, particle-particle hole medium polarization effects are incorporated in our nuclear medium cooling scenario in all processes in contrast with the minimal cooling paradigm which is disregarding these effects.

The proton branch of the MMU, $np\to ppe\bar\nu$, and the medium nucleon bremsstrahlung (MNB) reactions, $NN\to NN\nu\bar\nu$,   are also enhanced in the nuclear medium cooling scenario for $n\gsim n_0$ owing to the pion softening effect which is increasing with the density.
The MNB processes contribute to the emissivity less than the MMU processes, except for cases of very large gaps of one kind ($pp$ or $nn$).

Note \cite{Voskresensky:1986af,Senatorov:1987aa,Migdal:1990vm} that the contribution of the intermediate reaction states is the largest in the emissivity of the MMU processes for $n\gsim n_0$. This contribution is not incorporated at all in the MU emissivity used in the minimal cooling scheme. 
On the contrary, the correlation effects for $n<n_0$ result in a suppression of the MMU and MNB emissivities compared to MU and NB ones, cf. \cite{Blaschke:1995va,Hanhart:2000ae,Knoll:1995nz}.

In Refs. \cite{Blaschke:2004vq,Kolomeitsev:2004ff,Klahn:2006ir,Blaschke:2006gd} the so called direct Urca (DU) constraint has been introduced. 
The emissivity of this reaction, $n\to pe\bar{\nu}$, is $\sim 10^6$ times higher than that of the MU reactions computed following \cite{Friman:1978zq} via the free one-pion exchange, i.e. without inclusion of the pion polarization effects. 
The DU emissivity is also by 2-3 orders of magnitude higher than that of the MMU processes for massive stars. 
So, when the DU reaction gets switched on as a function of the CS mass the cooling curves drop down below the data points. 
Based on the assumption that the mass distribution of  those objects for which surface temperatures are measured is similar to the one extracted, e.g., from a population synthesis analysis and from  supernova simulations, the very efficient DU reaction should be forbidden in the majority of the former CS, see \cite{Klahn:2006ir,Blaschke:2006gd}.  
So, as a ``weak" DU constraint Ref.~\cite{Klahn:2006ir} suggested to use 
$M_{\rm DU}^{\rm weak}=1.35~M_{\odot}$, the statistical average value for masses of binary radio pulsars, since it is natural to assume that the stars with such a mass  should be presented in $T_s -t$ plane in sufficient number. 
As a ``strong" DU constraint Ref.~\cite{Klahn:2006ir} suggested to use 
$M_{\rm DU}^{\rm strong}=1.5~M_{\odot}$, since the relative number of  objects with $M>1.5~M_{\odot}$  should be rather small, as follows from the population synthesis analysis. 
The ``strong" constraint could be  weakened, if one exploited very large proton pairing gaps.

The influence of in-medium effects on the NS cooling was first demonstrated
in \cite{Schaab:1996gd} with various  EoS.
The nuclear medium cooling scenario which was systematically developed
further in \cite{Blaschke:2004vq,Grigorian:2005fn,Blaschke:2011gc}
exploiting both, the HHJ EoS (a causality-preserving modification of the APR EoS) and the 
for $n>4~n_0$ stiffer HDD EoS \cite{Blaschke:2013vma}. 
It provides a successful description of the whole set of known cooling data for NS with low magnetic fields.

The resulting cooling curves are rather insensitive to the values of the $1S_0$ $nn$ pairing gaps but sensitive to the choice of the $1S_0$ $pp$ gaps and $3P_2$ $nn$ gaps, since $1S_0$ $nn$ pairing gaps drop already for $n\gsim 0.6 - 0.8 n_0$, whereas $1S_0$ $pp$ gaps  are spreading  to $n\sim 2.5 - 4 n_0$, and $3P_2$ $nn$ gaps may spread to a higher density. 
Indeed,  the dense interior rather than the crust determines the total luminosity within our scenario. 
The gaps are sensitive to in-medium (nucleon-nucleon loop) effects, and are not well known due to exponential dependence on the $NN$ interaction amplitude  in the pairing channel.
Especially the values of the $3P_2$ neutron pairing gaps are poorly known. 
Estimates \cite{HGRR1970,Takatsuka:2004zq} produce  typical values of  
$\Delta({\rm 3P_2}) \sim 0.1 - 1$ MeV.
Accounting for medium-induced spin-orbit interaction, Ref.~\cite{Schwenk:2003bc}  computed a tiny value of $\Delta({ 3P_2}) \lsim 10$ keV, whereas Ref.~\cite{Khodel:2004nt} argued that $\Delta({ 3P_2})$ might be very large, $\gsim $MeV, if a very strong enhancement of the tensor $NN$ interaction in the particle-particle channel occurs, owing to the softening of pion modes.  
In our model an overall fit of the CS cooling data is obtained for a strongly suppressed value of the $3P_2$ neutron pairing gap, thus being in favour of  results of Ref.~\cite{Schwenk:2003bc}. 
The dependence of the cooling curves on the $3P_2$ neutron  and $1S_0$ proton paring gaps was studied within our scenario in  \cite{Grigorian:2005fn}. 
The successful description of all cooling data within our scenario, where many  in-medium effects are shown to be important while they are being disregarded in the minimal cooling scenario, demonstrates 
that the statement made within the latter scenario in some works, that an appropriate fit of existing  Cas A data  allows  to ``measure" the critical temperature of the $3P_2$ $nn$ pairing as $T_c \sim (5 - 9) \cdot 10^8$ K, is not justified.

Reference \cite{Blaschke:2011gc} has demonstrated an appropriate fit of the Cas A cooling curves with
results from ACIS-S instrument yielding a surface temperature decline of $\sim 3\%$ over 10 yrs.
For that  the lepton heat conductivity has been suppressed artificially by a
factor $\sim 0.3$ in the most favourable case compared to the result
of Ref.~\cite{Baiko:2001cj}, which has been exploited as optional value in our previous works. 
The nucleon contribution to the heat conductivity is suppressed in our case by medium effects compared to that used in Ref.~\cite{Baiko:2001cj}.
In a more recent work \cite{Blaschke:2013vma} we have used the result of Ref.~\cite{Shternin:2007ee} for the lepton heat conductivity which includes Landau damping (electron-electron hole polarization) effects. An appropriate fit of   $\sim 3\%$ cooling curve decline for the Cas A was then obtained
without applying any artificial suppression of the lepton heat conductivity and the mass of the object 
was predicted as $\simeq 1.5~M_{\odot}$.
The HHJ hadronic EoS \cite{Heiselberg:1999fe}, which has been exploited in our previous works, was stiffened  in \cite{Blaschke:2013vma} for $n \gsim 4n_0$ to comply with the constraint that the EoS
should allow for a maximum NS mass above the value $M=2.01\pm 0.04~M_\odot$ measured for
PSR J0348+0432 by \cite{Anotoniadis}, see also \cite{Demorest}.
However, the resulting EoS (labeled as HDD), which produces
$M_{\rm max}=2.06~M_{\odot}$,  still might be not sufficiently stiff, since
the existence of even more massive objects than
those known up to now \cite{Demorest,Anotoniadis} is not excluded.
Incorporating systematic light-curve differences, the
authors of  \cite{Romani} have estimated that the mass of the
black-widow pulsar PSR J1311-3430 should at least be $M>2.1~M_{\odot}$.
Furthermore, a deconfinement transition in the CS interior
would contradict these  mass measurements,  if one would use a  soft
hadronic EoS, cf. \cite{Alford:2006vz,Klahn,Klahn2}.
Therefore, the investigation of the possibility of hybrid stars requires a stiff hadronic EoS.
Also, recent radius determinations from timing
residuals suggest large CS radii (albeit still with large uncertainties) and
thus motivate the usage of a stiffer hadronic EoS \cite{Bogdanov:2012md}. 
Note that the authors of the recent paper \cite{Ho:2014pta}
have explained the Cas A ACIS-S data for the NS mass $M=1.44~M_{\odot}$
within the minimal cooling scenario using the stiff BSk21 EoS,  a large proton gap
and a moderate $3P_2$ neutron gap. 
Hottest and coldest objects, however, can hardly be explained appropriately within the same scenario.

Note also for completeness that the authors of Ref.~\cite{Posselt:2013xva}
indicated deficiencies in analysis that have led to  ACIS-S data and they have argued that  the decline extracted from the ACIS-S  graded mode data might be irrelevant thereby. 
However, the authors  who presented the  ACIS-S data, see \cite{Ho:2014pta}, continue to use their own data analysis, although mentioning results of other analyses. 
Due to  the presently existing uncertainty in the  interpretation of the Cas A data, when demonstrating our results we compare them with both the ACIS-S graded mode data  and the HRC-S data, in order to guide the eye. The latter data are compatible with cooling curves having a smaller decline.

In a previous contribution \cite{Grigorian:2015nva} we have
shown  preliminary calculations of the CS cooling  within our nuclear medium cooling
scenario exploiting the stiff DD2 EoS \cite{Typel:2009sy}.
However, in that work, being mostly interested in demonstrating results of a hybrid star model, we have not performed any  tuning of the parameters of the hadronic model that we exploited in \cite{Blaschke:2011gc}.

In this paper, we will be exploiting the stiff DD2 and the still stiffer excluded volume DD2vex EoS
first within the hadronic scenario. 
By tuning the proton pairing gaps and the effective pion gap we will construct appropriate  fits of the cooling data, see \cite{Grigorian:2015oaa}. 
Note that in absence of the pion condensation a strong pion softening for $n>n_0$ does not contradict to the stiffening of the EoS since the strength of the attraction in the pion self-energy is controlled by the zero-harmonics  of the spin Fermi liquid  Landau-Migdal parameters $g_0^{nn}$ and $g_0^{np}$, whereas the stiffening of the EoS is determined by the scalar parameters $f_0^{nn}$ and $f_0^{np}$.

The strong interaction in quark matter ($\alpha_s\neq 0$) permits the quark DU process \cite{Iwamoto}. First studies have shown that this result does not exclude the possibility that some CS might possess large quark matter cores. 
The suppression of the emissivity of the DU process may arise from the occurrence of diquark pairing and thus color superconductivity in the quark matter phase \cite{BKV,Page,BGV2001}. 
Ref.~\cite{Grigorian:2004jq} discussed a "2SC + X" phase as a possibility for having all quarks paired in two-flavor quark matter, where the density dependent X gap is of the order of 10 keV - 1 MeV.  
The presence of an X gap that decreases with increasing density allowed to appropriately fit the cooling data in a similar CS mass interval to that following from a purely hadronic model.  
Ref.~\cite{Grigorian:2004jq} exploited the HHJ EoS for hadronic matter matched with the NJL EoS for quark color superconducting matter. 
Thereby the  constraint $M>1.97~M_{\odot}$, which follows from recent measurements  of the masses of the pulsars PSR J1614-2230 \cite{Demorest} and PSR J0348-0432 \cite{Anotoniadis}   was not fulfilled. Thermal evolution of CS with crystalline and alternative gapless color-superconducting phases were studied in \cite{HessSedrakian}.  Refs.~\cite{Sedrakian:2013xgk,Grigorian:2015nva} demonstrated that Cas A data could be explained  within the  hybrid star scenario exploiting stiff hadronic EoS.

As an alternative to the purely hadronic scenario, in the given work we also discuss the possibility of the deconfinement phase transition from hadronic matter in the outer core to color superconducting quark matter in the inner core of most massive NS.  
A  stiff hadronic EoS, like DD2vex,  paves the way for exploring scenarios with a phase transition to
quark matter, see \cite{Alford:2006vz,Klahn,Klahn2} for recent examples, which despite a softening due to the phase transition meet the constraint of the $2M_{\odot}$ pulsar mass measurement. 
The quark matter EoS is described by a color superconducting two flavor NJL model in the
2SC phase with a stiffening due to a vector mean field.

The paper is organized as follows. 
In  section \ref{sec:EoS} we introduce the EoS studied in the given work and demonstrate corresponding mass-radius relations. 
In section \ref{sec:Cooling} we explain our cooling models first for hadronic stars and then for hybrid stars. Then in section \ref{sec:Results} we show results of calculations of the NS cooling first with the HDD, and then with the DD2 and DDvex EoS, and finally we discuss hybrid star cooling with the DDvex+QM EoS. 
Our conclusions are presented in section \ref{sec:Remarks}. 
In an Appendix we argue that to avoid severe inconsistencies in the treatment of various in-medium effects the minimal cooling scheme should be extended including particle-particle hole polarization loop effects. The most essential ones of these effects are included into consideration within our nuclear medium cooling scenario.

\section{\label{sec:EoS}Equation of state}

In our previous works \cite{Blaschke:2004vq,Grigorian:2005fn,Blaschke:2011gc} we
have exploited the HHJ ($\delta =0.2$) fit \cite{Heiselberg:1999fe} of the APR EoS \cite{Akmal:1998cf}.
While the latter EoS produces an appropriate maximum NS mass of $M=2.2~M_{\odot}$,
the HHJ EoS  fit introduces an additional parametric correction of the high-density
behaviour in order to avoid a causality breach that occurs with the APR EoS in that region.
This comes at the price of a lowering of the maximum NS mass to $M=1.94~M_{\odot}$,
below the $1\sigma$ limit for the recently measured mass of the pulsar PSR J0348-0432 \cite{Anotoniadis}.
Therefore, we have recently modified this EoS by invoking an excluded volume for nucleons
\cite{Blaschke:2013vma}.
The so constructed HDD EoS stiffens for higher baryon density $n$ resulting in
an increase of the maximum NS mass up to the value $M_{\rm max}\simeq 2.06~M_{\odot}$. 
The maximum central density is $n^{\rm max}_{\rm cen}\simeq 7.7~n_0$. 
The DU processes start for $M>M_{\rm DU}\simeq 1.88~M_{\odot}$ what corresponds to a 
central density of $n_{\rm cen}\simeq 5~n_0$. 
The radius of the $1.4 M_{\odot}$ NS is about $12$ km, agreeing with the analysis of NS data in \cite{Lattimer} and with a recent analysis using the cooling tail method \cite{Nattila:2015jra}, see also
Ref.~\cite{Klochkov} and the contribution to this volume \cite{Suleimanov:2015fwa}.

However a  radius determination for the nearest millisecond pulsar PSR J0437-4715 
\cite{Bogdanov:2012md} supports (at $2\sigma$ confidence) the value $\geq 13$ km  in the mass segment
between 1.5 and 1.8 $M_\odot$, see also \cite{Bogdanov:2015tua}. 
A  stiffer EoS is needed to support such a large CS radii.
The density dependent relativistic mean-field EoS of Ref.~\cite{Typel:1999yq} with the  DD2 parametrization \cite{Typel:2009sy} meets this requirement.
Moreover, the DD2 EoS fulfils  standard constraints for
symmetric nuclear matter around the saturation density and from nuclear structure.
The density dependent symmetry energy  agrees
with the  constraint  \cite{Danielewicz:2013upa} and with ab-initio
calculations for pure neutron matter \cite{Hebeler:2010jx}.
The DU reaction threshold is not reached within the DD2 EoS. The maximum NS mass
is $M_{\rm max}=2.43~M_{\odot}$  what corresponds to a central density of $n_{\rm cen}\simeq 5.1~n_0$.
However, due to the stiffness of the DD2 EoS, it does not fulfil the  "flow constraint" \cite{Danielewicz:2002pu} for densities above $2n_0$.
This is the price to be paid for the possibility to  increase the maximum mass and the radii of NS.

As we have mentioned, some measurements, although performed with large experimental error bars and with additional theoretical uncertainties, indicate still higher masses of the objects, see Fig.~4 of 
\cite{Watts:2014tja}. 
For example, Ref.~\cite{Romani} obtain the mass of the  $\gamma$-ray  pulsar PSR J1311-3430 as $2.68\pm 0.14~M_{\odot}$ for the basic light curve model and to $2.15\pm 0.11~M_{\odot}$ when a cool spot is assumed.
Therefore, we perform in this work also calculations with the still stiffer DD2vex EoS.

The DD2vex EoS belongs to a class of DD2 EoS that has been developed in order to account for additional repulsive $NN$ interactions at supersaturation densities that arise from the compositeness of baryons. 
Quark and gluon substructure effects in the NN interaction become operative at densities where the nucleon wave functions start overlapping. For a discussion of quark Pauli blocking and multi-pomeron exchange contributions to a universal many-body repulsion see, e.g., the contribution by Yamamoto et al. to this volume \cite{Yamamoto:2015lwa} and references therein.
The requirement of the antisymmetrization of the two-nucleon wave function with respect to quark exchange leads to the impenetrability (hard core) of the nucleons, resulting in an excluded volume effect. 
In Ref.~\cite{Ropke:1986qs} it has been demonstrated on the basis of a nonrelativistic potential model for the hadron structure that the quark Pauli blocking effect results in a repulsive density dependent interaction that is well compatible with the repulsive part of Skyrme-type interactions.
Recently it has been shown within an exploratory calculation \cite{BGR:2016}
that the onset of partial chiral symmetry restoration, as triggered by the melting of the chiral condensate due to increasing phase space occupation, leads to a strong enhancement of this excluded volume effect.
In order to capture these fundamental symmetry aspects in the EoS for composite hadrons at high densities we will employ here a modern excluded volume prescription that is thermodynamically consistent and does not affect, by construction, the EoS at subsaturation and saturation density  where the EoS parameters are adjusted to nuclear structure data. 
For the very details of this new excluded volume prescription, see the contribution by S. Typel to this volume \cite{Typel:2016srf}. 
The DD2vex family of EoS \cite{Typel:1999yq} has been used as a basis for a Bayesian M-R analysis in order to extract the most probable EoS behaviour at high densities \cite{Ayriyan:2015kit}, 
Applying the constraint for the nearest millisecond pulsar PSR J0437-4715 \cite{Bogdanov:2012md,Bogdanov:2015tua} which yields rather large radii $\sim 15$ km, in this analysis  
the stiffest possible hadronic EoS supporting NS with large radii get selected as the most probable ones.
This large radius constraint has a preference over other results for the radii since its analysis is not flawed by the uncertainties due to unknown atmospheric composition \cite{Servillat:2012df}
or controversial interpretation of data from X-ray bursters, see the corresponding contributions 
to this volume \cite{Suleimanov:2015fwa,Lamb-Miller:2016}.
One of those stiffest possible EoS is DD2vex, which we will exploit in the given work,  with a van-der-Waals excluded volume parameter $v=16\pi r_N^3/3 = 4.0$ fm$^3$ that corresponds to a nucleon radius of $r_N\simeq  0.6$ fm. The causality breaches for  $M>M_{\rm max}=2.66~M_{\odot}$  that corresponds to the central density $n_{\rm cen}\simeq 2.55~n_0$. 
The DU  processes occur for $M>M_{\rm DU}=2.51~M_{\odot}$.

Besides a variation of the hadronic excluded volume parameter $v$, in \cite{Ayriyan:2015kit} 
also the transition to quark matter (QM) was considered within the two-phase approach, where a simple Maxwell construction has been performed to a quark matter EoS of NJL type with higher order quark interactions (hNJL)  \cite{Benic:2014iaa} of varying strength parameter $\eta_4$ in the isoscalar vector interaction channel.  
The maximum CS mass with the DD2vex-QM EoS for $\eta_4=5.0$ reaches $2.19~M_{\odot}$ at 
$n^{\rm max}_{\rm cen}=5.5~n_0$.

\begin{figure}[!thb]
   \includegraphics[width=0.57\textwidth]{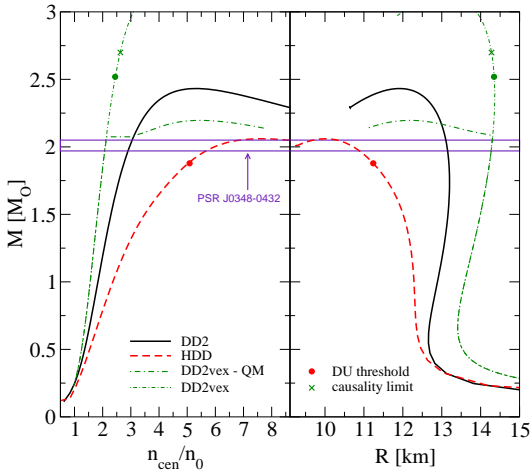}
     \caption{Mass vs. central baryon density (left) and vs. radius
     (right) for the HDD  (dash lines), a stiffer DD2   (solid lines) and a still stiffer DD2vex hadronic EoS (dash-dotted lines). Bold dash-dotted continuation is due to the quark (QM) part of the EoS. The crosses for the DD2vex hadronic EoS indicate violation of the causality. The band shows the experimental data on the measured mass for the pulsar PSR J0348-0432. The squares indicate the DU thresholds for the hadronic EoS. }
   \label{Fig:MR}
 \end{figure}

In Fig.~\ref{Fig:MR} we show the solutions of the TOV equations  $M$ vs. $n_{\rm cen}$ and $M$ vs. $R$ for the HDD,  DD2 and the DD2vex EoS with the rather conservative choice of $v=4$ fm$^3$ without and with a phase transition to hNJL quark matter with $\eta_4=5.0$, DD2vex-QM, see also \cite{Benic:2014jia}. The dependencies of CS masses on central baryon density $n_{\rm cen}$  are shown in the left panel  for the HDD (dash curve), DD2 (solid curve) and DD2vex (dash-dotted curve) hadronic EoS. 
In the latter case above a critical  density (see the cross at the dash-dotted curve) the sound velocity exceeds the velocity of the light and the causality breaches. 
Thus DD2vex EoS should not be used above this density. 
The DU thresholds for hadronic EoS are shown by the squares.
We see that for a fixed NS mass the stiffening of the hadronic EoS leads to a redistribution of the density profile in the NS interior so that the central densities get lowered.
As a consequence, a slower cooling is expected for stars with the DD2vex EoS, when compared with stars of the same mass described by the DD2 and especially the HDD EoS, provided other ingredients of the model are the same. 
The bold dash-dotted continuation of the dash-dotted curve shows the transition to the quark matter for the DD2vex-QM model explained above. The quark matter core appears for $M>2.08~M_{\odot}$ ($n_{\rm cen}> 2.1~n_0$).
The right panel of Fig.~\ref{Fig:MR} demonstrates the mass-radius relation. The stiffest DD2vex  EoS produces a larger radius than DD2 and the DD2 EoS yields a larger radius than the softer HDD EoS. 
For $M=1.5~M_{\odot}$ we have  $R_{\rm DD2vex}\simeq 14$ km, $R_{\rm DD2}\simeq 13$ km, $R_{\rm HDD}\simeq 12$ km. 
Thus appropriate mass-radius measurements could allow to put constraints on the stiffness of the EoS.

\section{\label{sec:Cooling}Cooling }

\subsection{\label{ssec:hadcool}Hadronic Stars}

In our previous works
\cite{Blaschke:2004vq,Grigorian:2005fn,Blaschke:2011gc,Blaschke:2013vma}
we have demonstrated that the cooling history is most sensitive to the
efficiency of the MMU process controlled by the
density dependence of the effective pion gap shown in Fig.~1 of
\cite{Blaschke:2004vq} and to the value and the density dependence
of the $1S_0$ $pp$ pairing gap.
Generally speaking the results are very sensitive to the values and density dependence of the $3P_2$ 
$nn$ pairing gap, as was demonstrated in Ref.~\cite{Grigorian:2005fn} by variation of this gap in broad limits.
In our works we follow the analysis of \cite{Schwenk:2003bc}, where this gap turns out
to be negligibly small. 
Then we get the best overall fit of the cooling data. 
The influence of the crust on the cooling is rather minor, especially in our model, where the cooling of the core is enhanced by medium effects on MMU and MNB processes. 
Our results are also rather insensitive to the treatment of the $1S_0$ $nn$ pairing gap, as we have demonstrated in \cite{Grigorian:2005fn} by its variation, since it is not spread from the crust into
 the interior region. 
Thereby this gap is taken the same as in our previous works, cf. Fig.~5 in \cite{Blaschke:2004vq}.
In \cite{Blaschke:2011gc,Blaschke:2013vma} exploiting the HHJ and the HDD EoS we have also demonstrated that the decline of the cooling curve describing the evolution of the Cas A surface temperature is sensitive to the value of the heat conductivity in the CS interior region.
In \cite{Blaschke:2013vma} we have demonstrated that with an appropriate choice of the proton pairing gap (following model I) we are able to fit  the $\gsim 2.5 \%$ decline as was shown by ACIS-S data on 
Cas A using the same lepton heat conductivity as in \cite{Shternin:2007ee}. 
With the gaps given by model II and using the same lepton heat conductivity we match a $\lsim$ 1-2$\%$ decline, cf. \cite{Blaschke:2013vma}.

In the given work the neutrino emissivities, specific heat,
crust properties, etc. are taken from our earlier works performed on the basis of the HHJ EoS \cite{Blaschke:2011gc} and the HDD EoS \cite{Blaschke:2013vma} for hadronic matter.
The heat conductivity is the same as in \cite{Blaschke:2013vma}.
The best fit of Cas A ACIS-S data with the HDD EoS was obtained in \cite{Blaschke:2013vma}
with the same effective pion gap and the same $1S_0$ $pp$ pairing gap of the model I
as in our previous works \cite{Blaschke:2004vq,Grigorian:2005fn,Blaschke:2011gc}.
Now when we are exploiting the still stiffer DD2 and DD2vex EoS, we
will additionally tune the $pp$ pairing gap and the pion effective gap in order to get the best fit of the cooling data, while retaining all other values the same as in \cite{Blaschke:2013vma}.

 \begin{figure}[!thb]
   \centering
   \includegraphics[width=0.53\textwidth]{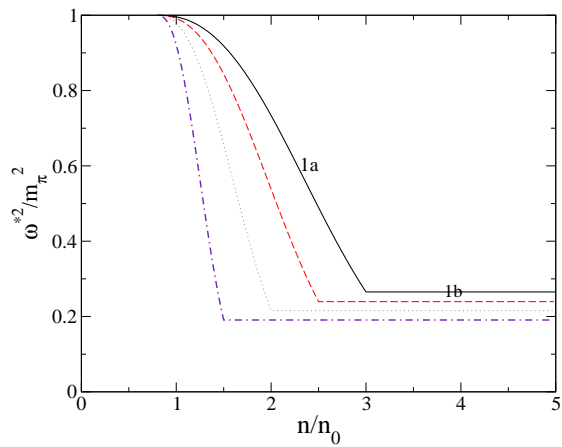}
     \caption{Square of the effective pion gap as a function of the density without pion
     condensation (curves 1a+1b), $m_{\pi}$ is the pion mass, but with a pion softening saturated above a critical value. The solid line corresponds to
     the same parametrization as in our previous work \cite{Blaschke:2004vq}, the other lines demonstrate a stronger pion  softening effect, for details see text.}
   \label{Fig:omegatilde}
 \end{figure}

The density dependence of the square of the effective pion gap ${\omega}^{*\,2} (n)$
that we exploit in the given work is shown in Fig.~\ref{Fig:omegatilde}.
To be specific and not to complicate the considerations we consider the case when the pion softening is saturated above a critical density $n_c^{\pi}$ and pion condensation does not occur.
The solid curve 1a+1b is precisely the same as  in Fig.~1 of \cite{Blaschke:2004vq}, demonstrating saturation of the pion softening for $n>n_c^{\pi}=3~n_0$. 
The dashed, dotted and dash-dotted lines show a stronger pion softening effect with a saturation for $n>n_c^{\pi}=2.5~n_0$, $n>n_c^{\pi}=2~n_0$, and $n>n_c^{\pi}=1.5~n_0$, respectively.
Note that variational calculations of Ref.~\cite{Akmal:1998cf}
 show that pion condensation in NS matter may appear already for $n>n_c^{\pi}\simeq 1.3~n_0$, thus 
 being in favor of a still steeper $\omega^*(n)$ dependence. 
Just in order to be as conservative as possible we continue to exploit a  weaker pion softening in our calculations. 
\begin{figure}[!t]
   \includegraphics[width=0.53\textwidth]{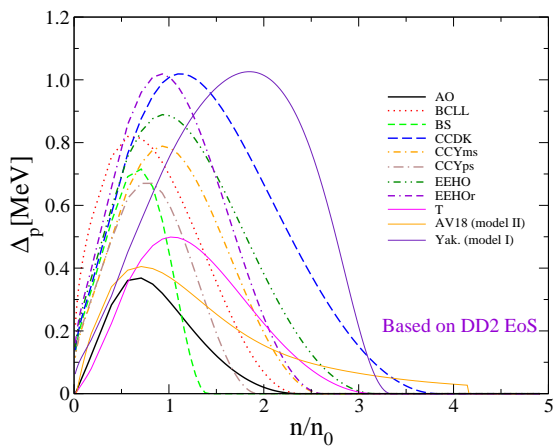}
     \caption{$1S_0$ $pp$ gaps as  functions of the baryon  density for zero temperature.
     The abbreviations in the legend correspond to those used in Ref.~\cite{Ho:2014pta}.
     The gaps labeled as ``Yak" and ``AV18" are those (models I and II, respectively)
     exploited in our previous works
     \cite{Blaschke:2004vq,Grigorian:2005fn,Blaschke:2011gc,Blaschke:2013vma}.}
   \label{Fig:gaps}
 \end{figure}

We stress that the pion softening appears  only for $n>n_{c1}$ choosen to be $0.8~n_0$.
Only then there appears a minimum in the effective pion gap $\omega^*(k)$ for $k=k_0 (n)\neq 0$, 
$\omega^*(k_0)\equiv \omega^* (n)$. 
For $n<n_{c1}$ there is no reason for the pion softening thereby. 
As a result, the emissivity of the MNB process proves to be several times suppressed for $n<n_{c1}$ as compared to the values computed within the FOPE model because correlation effects are  incorporated in the particle-particle channel, in the same manner as they have been taken into account in \cite{Blaschke:1995va,Hanhart:2000ae}. 
The MMU emissivity is less affected by these correlation effects for $n<n_{c1}$, see \cite{Knoll:1995nz}. 
Sometimes these important points are overlooked by authors who continue to exploit emissivities 
for all two-nucleon processes which are at $n<n_c1 \sim 0.5-0.8~n_0$ several times  suppressed compared to those computed  in the FOPE based model \cite{Friman:1978zq} also  for $n>n_0$, 
where the pion softening effect should be dominant and leads to an enhancement compared to the FOPE case.

The $1S_0$ $pp$ pairing gaps for zero temperature exploited by different authors are
shown in Fig.~\ref{Fig:gaps}, now for the DD2 EoS. 
In our nuclear medium cooling scenario we use the temperature dependence of the gaps    
taken as in \cite{Levenfish} and the fermion phase space in MMU, MNB and DU processes and specific heat is corrected by the corresponding $R$ factors.
We use the parametrization of the zero-temperature $pp$ pairing gaps, 
$\Delta_p (p_{{\rm F},p})$, from \cite{Ho:2014pta}, Eq. (2);  $p_{{\rm F},i}$ denotes the
Fermi momentum of the species $i$.
The parameters are taken to fit the gaps computed in various publications.
The abbreviations of the curves in Fig.~\ref{Fig:gaps} are taken over from Table II of \cite{Ho:2014pta}.
Two additional curves labeled as ``Yak" and ``AV18" correspond to the models I and II, respectively, exploited in our previous works \cite{Blaschke:2004vq,Grigorian:2005fn,Blaschke:2011gc} for the HHJ
EoS and in \cite{Blaschke:2013vma} for the HDD EoS.

With these gaps and the $\omega^*$ parametrizations we compute the NS cooling history.

The density range $n\sim 0.5 - 0.8~n_0$  is the boundary between the NS interior and the inner crust. 
The latter is constructed of a pasta phase \cite{Maruyama:2005vb}. 
Up to now the pasta phase is not included in the NS cooling codes.  
Then at very low densities there is the outer crust and the envelope. 
The influence of the crust on the cooling and heat transport is rather minor,  because of its rather low mass content. 
In our scenario, where the cooling of the interior is strongly enhanced owing to the pion softening effect the role of the crust is still diminished.
Thus, the temperature changes only slightly in the region from the crust to the envelope. 
Due to the said above, and in absence of a study of the influence of pasta structures on the NS cooling, in a simplifying consideration we exploit the same EoS till the low densities.

In our scenario we use a fixed relation between the surface ($T_s$) and internal ($T_{in}$) temperatures  
(see the curve ``our fit" in Fig.~4 of \cite{Blaschke:2004vq}) within the band computed in \cite{Yakovlev:2003ed}  demonstrating a similar trend as the well known  ``Tsuruta law" \cite{Tsuruta:1979}. 
Our $T_s - T_{in}$ relation qualitatively takes into account that the hotter and younger objects may have less heavy  elements in  the atmosphere than the colder and older ones. A dependence of the cooling history on the choice  of the relation between  $T_s$ and $T_{in}$ was demonstrated in \cite{Blaschke:2004vq,Grigorian:2005fn}.

\subsection{\label{ssec:hybcool}Hybrid stars}
In order to describe the possibility of hybrid star configurations we  adopt the cooling of the quark core
following the lines of Ref.~\cite{Grigorian:2004jq}. 
For a hybrid star with a quark matter core in the 2SC phase, stable configurations with masses  above 
1.25 $M_{\odot}$ have been obtained, when a Gaussian formfactor regularization has been used \cite{Grigorian:2004jq}. 
This phase has one unpaired color of quarks (say blue) for which the very effective quark DU process
works and leads to a too fast cooling of the hybrid star in disagreement with the data.
Therefore in \cite{Grigorian:2004jq} we have suggested the presence of a weak pairing $X$-channel which could lead to a small residual pairing of the hitherto unpaired blue quarks. 
Other quark reaction channels are assumed to be blocked by the pairing with large gaps.
We have called the resulting gap $\Delta_X$ and could show that for a density dependent ansatz
$\Delta_X(\mu) = \Delta_c \mbox{exp}[-\alpha (\mu -\mu_c)/\mu_c]$
with $\mu$ being the quark chemical potential, $\mu_c = 330$ MeV, $\alpha =10$ and $\Delta_c = 1.0$ MeV an acceptable cooling phenomenology can be obtained \cite{Grigorian:2015nva}.

Other pairing schemes have been invoked which would obey the above requirement that all quarks have
to be paired, but the smallest pairing gap has not to exceed a size of $\sim 0.5 \dots 1.0$ MeV.
Most prominent in this context is the color-spin-locking (CSL) phase \cite{Schmitt:2003xq,Pang:2010wk} which has been evaluated in detail in the isotropic case for the local \cite{Aguilera:2005tg} and nonlocal  NJL model \cite{Aguilera:2006cj}. 
The neutrino emissivity and bulk viscosity of such a phase have been discussed in 
Ref.~\cite{Blaschke:2007bv,Blaschke:2008gd,Blaschke:2009je}.
However, no hybrid star cooling simulations have yet been performed with such a scenario.

\begin{figure}[!thb]
   \centering
   \includegraphics[width=0.75\textwidth,angle=0]{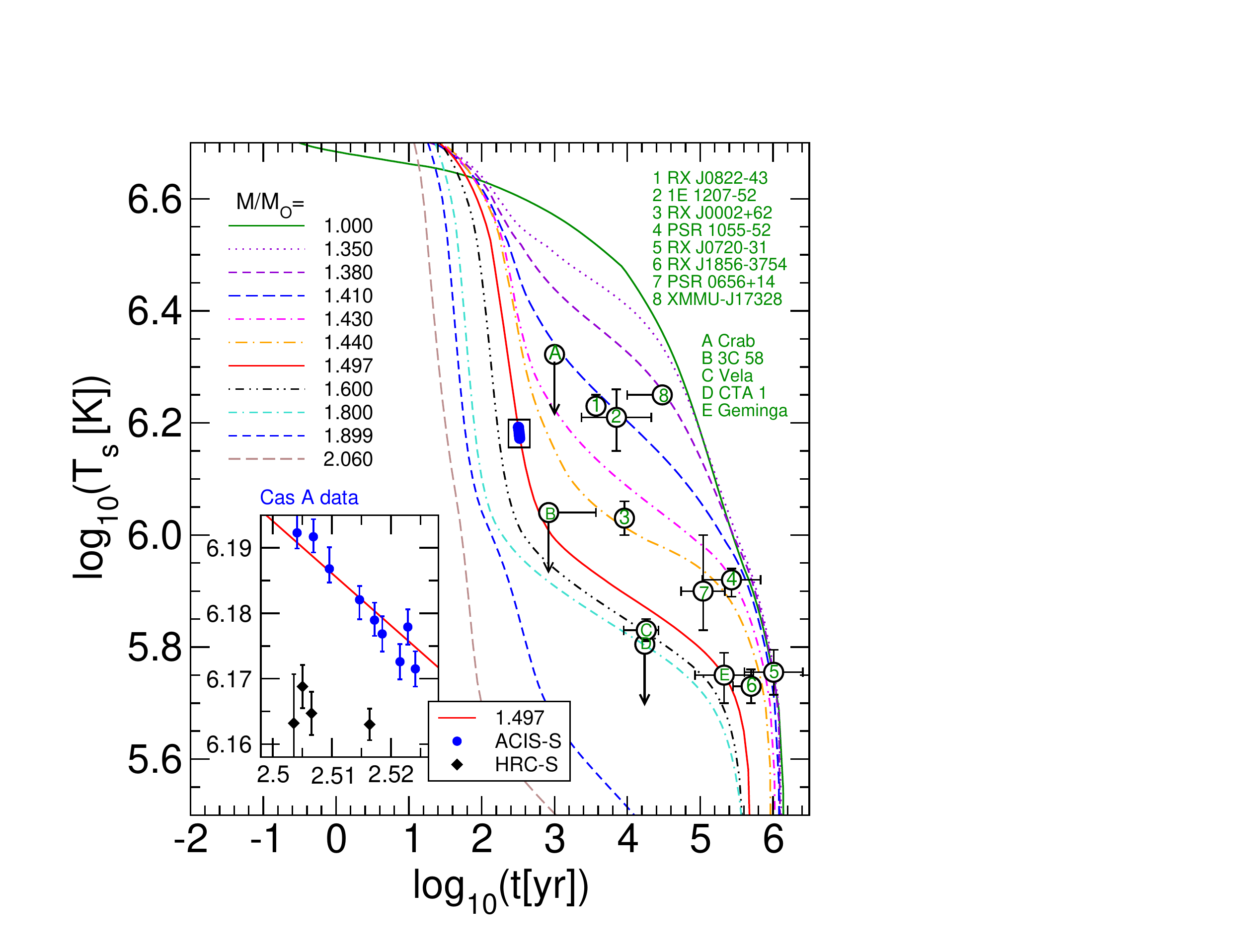}
      \caption{Cooling curves for a NS sequence according to the  hadronic HDD EoS;
      $T_s$ is the redshifted surface temperature, $t$ is the NS age.
      The effective pion gap is given by the solid curve 1a+1b in Fig.~\ref{Fig:omegatilde}, $n_c^{\pi}=3n_0$.
      The $1S_0$ $pp$ pairing gap corresponds to model I.
      The  mass range is shown in the legend.
      Comparison with Cas A ACIS-S and HRC-S data is shown in the inset. Cooling ACIS-S data for Cas A
      are explained with a NS mass of $M=1.497~M_\odot$. }
   \label{Fig:Cool1}
 \end{figure}

 \begin{figure}[!thb]
   \centering
   \includegraphics[width=0.75\textwidth,angle=0]{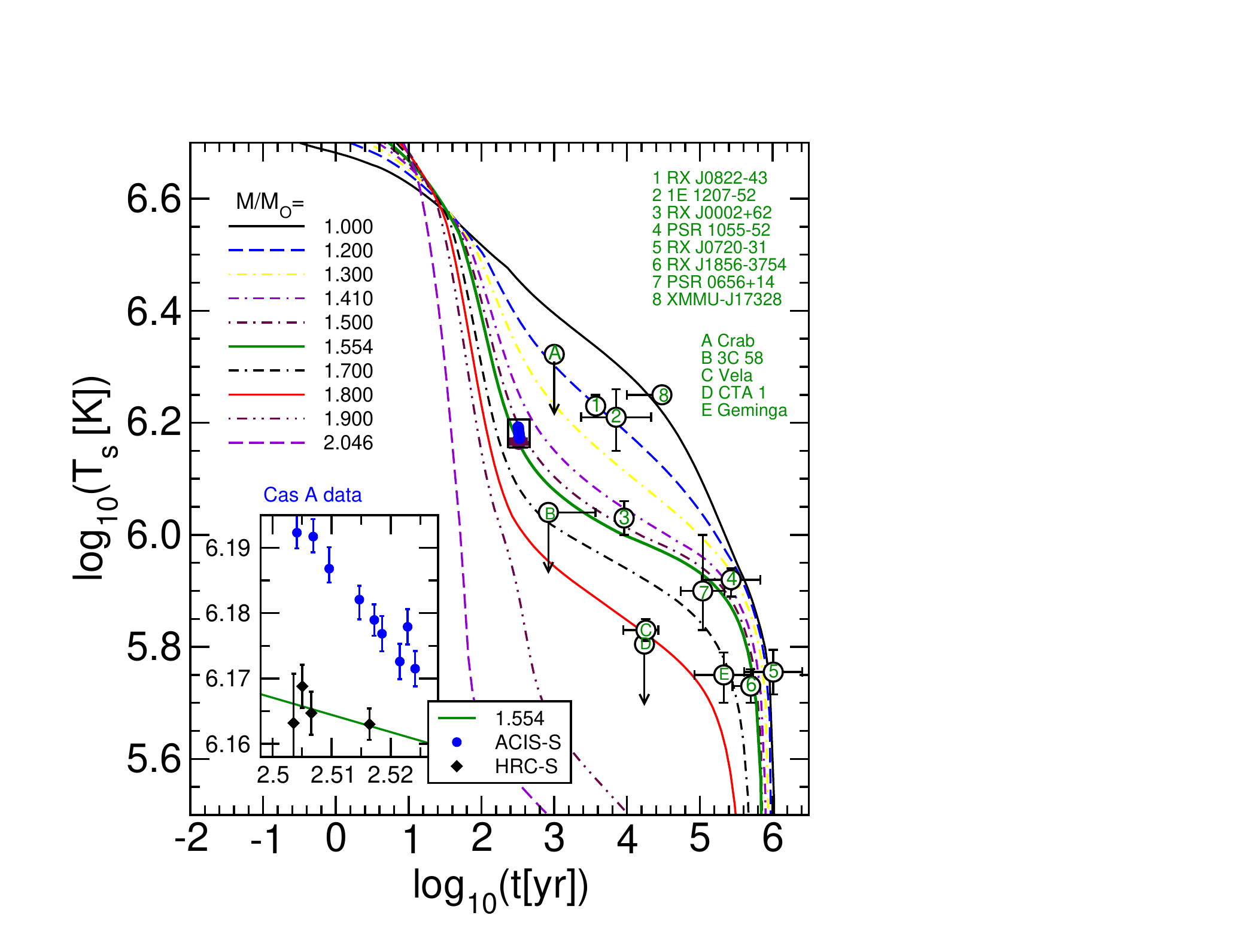}
      \caption{Cooling curves for a NS sequence according to the  hadronic HDD EoS;
      $T_s$ is the redshifted surface temperature, $t$ is the NS age.
      The effective pion gap is given by the solid curve 1a+1b in Fig.~\ref{Fig:omegatilde}, 
      $n_c^{\pi}=3~n_0$.
      The $1S_0$ $pp$ pairing gap corresponds to model II.
      The  mass range is shown in the legend.
       Comparison with Cas A ACIS-S and HRC-S data is shown in the inset. Cooling HRC-S data for Cas A  
       are explained with a NS mass of $M=1.554~M_\odot$.}
   \label{Fig:Cool1a}
 \end{figure}

\section{\label{sec:Results}Numerical results}

\subsection{NS cooling with the HDD EoS}

In Fig.~\ref{Fig:Cool1} we demonstrate the NS cooling history
computed with the HDD EoS in  Ref.~\cite{Blaschke:2013vma} using
model I for the $pp$ pairing gaps (see the gap ``Yak" in Fig.~\ref{Fig:gaps})
and with the effective pion gap given by the solid 1a+1b line  in Fig.~\ref{Fig:omegatilde}, for $n_c^{\pi}=3~n_0$. 
This model presents a natural generalization of the model shown in Fig.~12 in our work 
\cite{Blaschke:2004vq}.
As we see, the Cas~A ACIS-S data (with $\sim 2.5\%$ temperature decline per decade) are described by 
a NS with the mass $M_{\rm CasA} \simeq 1.497~M_{\odot}$.
A slight change of the value $M_{\rm CasA}$ compared to the value $1.541~M_{\odot}$ found in
\cite{Blaschke:2013vma} is due to unessential modifications of the parametrization in the present work. 
The whole range of data is covered by the cooling curves in the mass interval from 1.38 to $1.8-1.88  M_{\odot}$ ($1.88  M_{\odot}$ is the threshold value for the DU process). 
The Crab pulsar is described by a NS with the mass $\geq 1.4 M_{\odot}$ and Vela by 1.6 $M_{\odot}$. The XMMU source has a mass of $1.38-1.4 M_{\odot}$. 
All  sources, except the most rapidly cooling objects ``C and D'',  are covered by stars with masses from 1.38 to 1.5 $M_{\odot}$. 
The stars with $M>1.4~M_{\odot}$  cool down very rapidly since the $pp$ gap in the model I drops  sharply  to zero at densities corresponding to central densities of stars with $M> 1.4~M_{\odot}$. 
A high absolute value of the $pp$ gap in model I allows to explain a steep decline of the cooling curve for Cas~A compatible with the ACIS-S data.

In Fig.~\ref{Fig:Cool1a} we show the NS cooling history
computed with the HDD EoS in  Ref.~\cite{Blaschke:2013vma} using
model II for the $pp$ pairing gaps (see the gap ``AV18'' in Fig.~\ref{Fig:gaps})
and with the effective pion gap given by the solid 1a+1b line  in Fig.~\ref{Fig:omegatilde}, 
for $n_c^{\pi}=3~n_0$.
As we see, the Cas~A HRC-S data (with $\sim 0.7\%$ decline) are described by a NS with the mass $M_{\rm CasA} \simeq 1.554~M_{\odot}$. 
The Crab pulsar within this model has the mass $\geq 1.2~M_{\odot}$ and Vela has $M\simeq 1.8~M_{\odot}$.
The whole range of data is covered by cooling curves in the mass interval from 1.0 to 
$(1.8 - 1.88)~M_{\odot}$. 
The XMMU source has the mass $1-1.15 M_{\odot}$.  
All  sources, except  the most rapidly cooling objects ``C and D'',  are covered by stars with masses from 1 to 1.7 $M_{\odot}$. 
This model presents a natural generalization of the model shown in Fig.~20 in our work \cite{Blaschke:2004vq}. 
The latter model has passed the Log N-Log S test examined in \cite{Popov:2004ey}.  
A  broad distribution of NS masses from 1 to 2 $M_{\odot}$  follows from the neutrino-driven explosion models, cf. Fig.~6 in \cite{Ugliano:2012kq}. 
A smooth mass distribution that we obtain in this  model is controlled by a smooth density dependence of the $pp$ gap in model II. 
Due to a rather small value of the $pp$ gap  the decline of the curve describing Cas~A is substantially smaller than in the case of model I.

So, both the models (I and II) describe rather appropriately  the available $T_s (t)$ data. 
The Cas~A mass has a reasonable value $\sim 1.5~M_{\odot}$ in these models. 
If the decline for Cas~A were more carefully measured, one could decide,  which model is better.

Now, consider a cooling history exploiting a stiffer EoS bearing in mind that in the future may appear observations of more massive CS with larger radii, than those described by the HDD EoS.

\subsection{NS cooling with the DD2 EoS}
In our scenario, where we use a fixed $T_s - T_{in}$ relation between surface and internal temperature (see ``our fit" curve in Fig.~4 of \cite{Blaschke:2004vq}) it is a rule that if the hot (slowly cooling) objects
are described as more massive stars, then also the intermediate and rapidly cooling objects are more massive. 
Therefore, in order to achieve a description where the values for the masses of all cooling objects, hot as well as cold ones, lie in the relevant range of typical neutron star masses, we select parameters of the models such that in all relevant cases the hottest objects, like XMMU,  have masses below $1.5 M_{\odot}$.

As the hadronic DD2 EoS is stiffer at high densities compared to the HHJ and the HDD EoS, it produces a smaller central density for the star of the given mass than the latter EoS.
Therefore it leads to a weaker cooling activity, provided the same inputs are used for the effective pion gap ${\omega}^{*} (n)$, the pairing gaps and the other model ingredients. 
As a result,  under the above assumptions the neutron star in Cas~A should be described with the DD2 EoS as a more massive object than with the HHJ and HDD EoS. 
Ref.~\cite{Grigorian:2015nva} demonstrated  that when changing the EoS a description of all cooling data is possible even without changing any of the formerly adjusted cooling inputs except a tuning of the heat conductivity (in line with the strategy applied before in Ref.~\cite{Blaschke:2011gc}). 
However, in that case the NS mass that is required to fit the Cas~A cooling data with the stiff DD2 EoS
then amounts to $M=2.426~M_\odot$ for the model II of the pairing exploited there. 
Below we show that the latter quantity, being unrealistically high, can be diminished, if one performs a readjustment of the parameters of the model.

 \begin{figure}[!t]
   \centering
   \includegraphics[width=0.75\textwidth,angle=0]{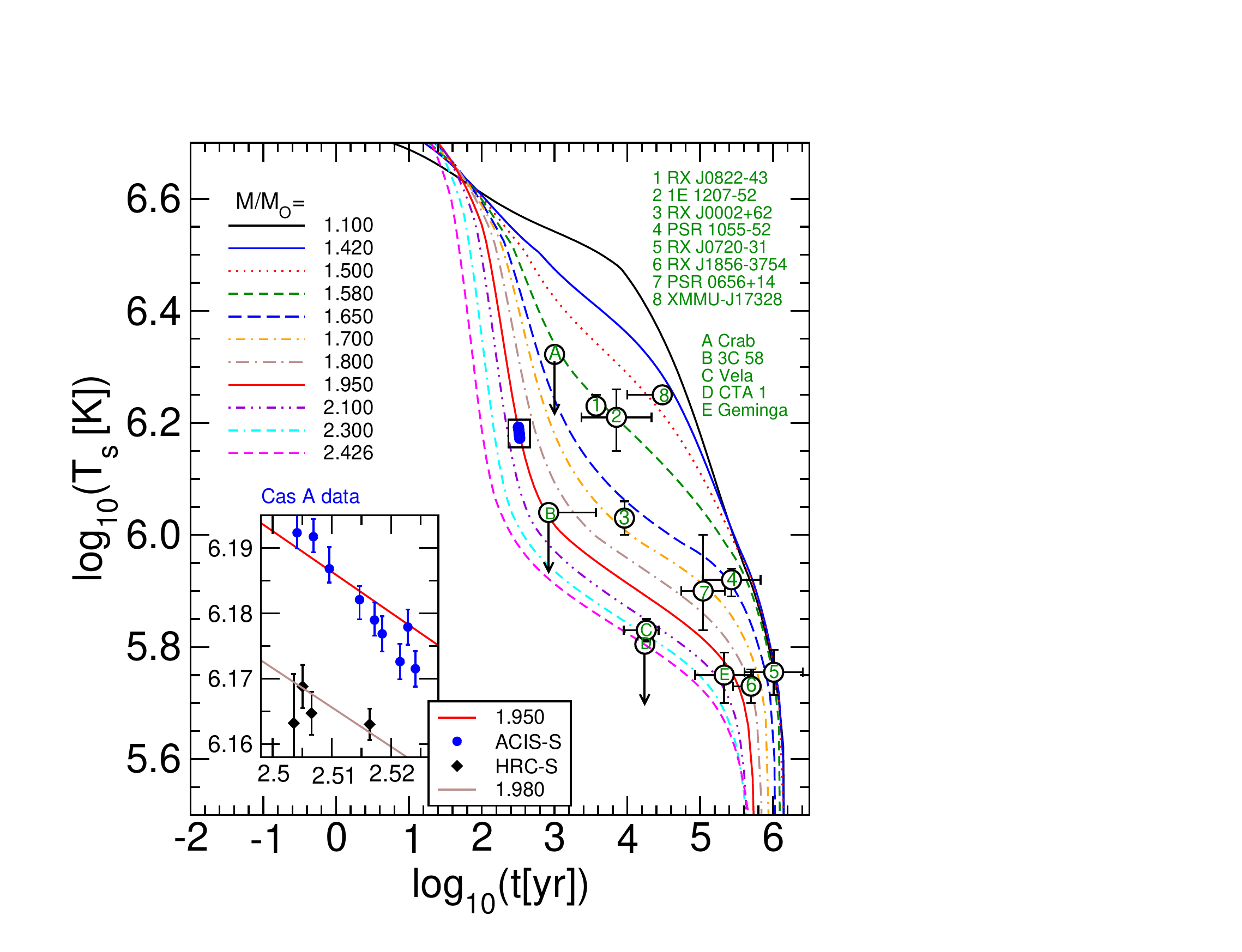}
      \caption{Cooling curves for a NS sequence according to the  hadronic
      DD2 EoS;
      $T_s$ is the redshifted surface temperature, $t$ is the NS age. The
      effective pion gap is given by solid curve 1a+1b in Fig.~\ref{Fig:omegatilde}, $n_c^{\pi}=3~n_0$.
      The $1S_0$ $pp$ pairing gap corresponds to model EEHOr.
      The  mass range is shown in the legend.
       Comparison with Cas~A ACIS-S and HRC-S data is shown in the inset. 
       The cooling data for Cas A from the ACIS-S instrument are explained by a NS mass 
       $M=1.950~M_\odot$, those of the HRC-S instrument by $M=1.980~M_\odot$.}
   \label{Fig:Cool2}
 \end{figure}

   \begin{figure}[!htb]
   \centering
   \includegraphics[width=0.75\textwidth,angle=0]{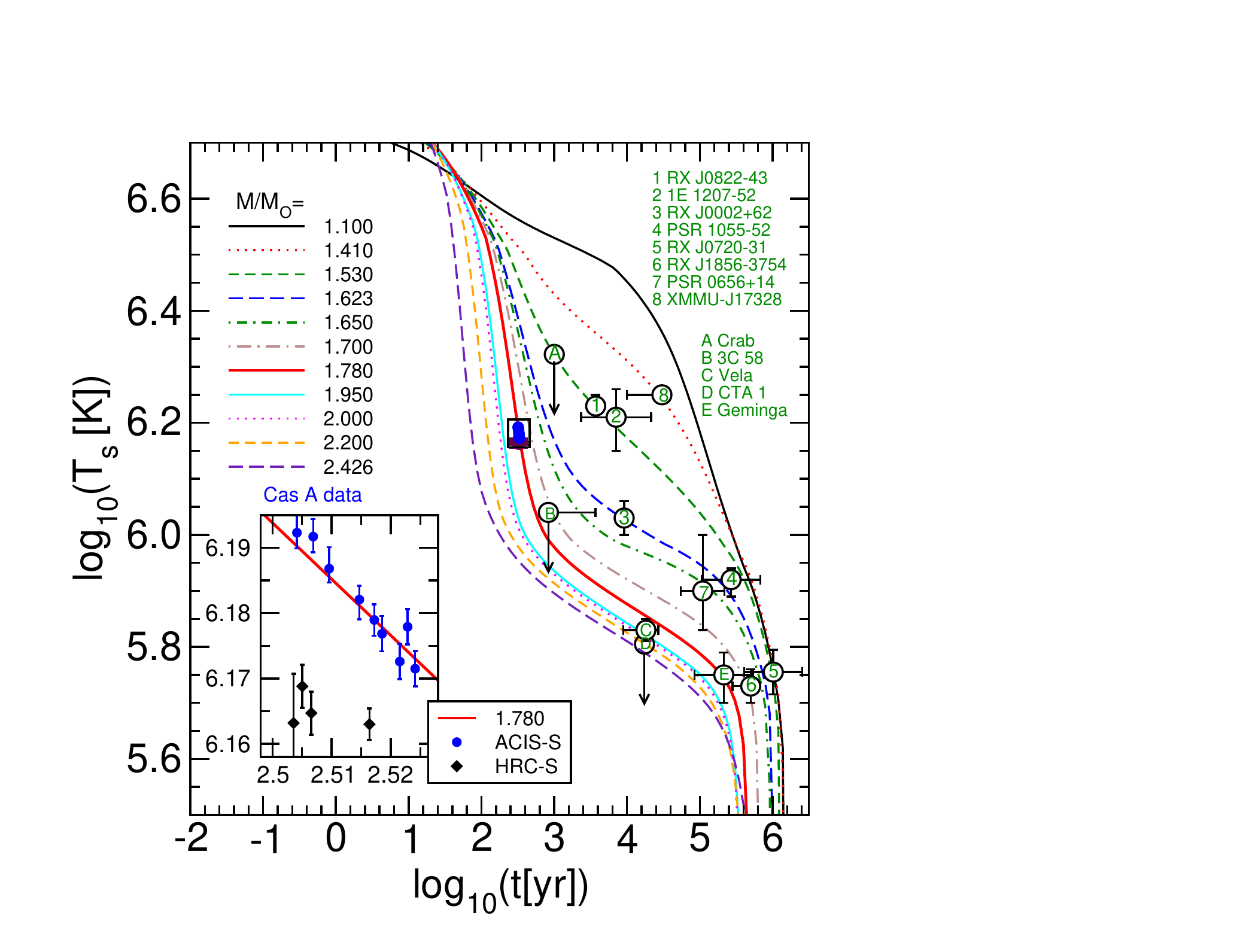}
      \caption{Same as Fig.~\ref{Fig:Cool2} for $n_c^{\pi}=2.5~n_0$.
Cas A cooling data from ACIS-S are explained with a NS of $M=1.780~M_\odot$. }
   \label{Fig:Cool3}
 \end{figure}

\begin{figure}[!htb]
   \centering
   \includegraphics[width=0.75\textwidth,angle=0]{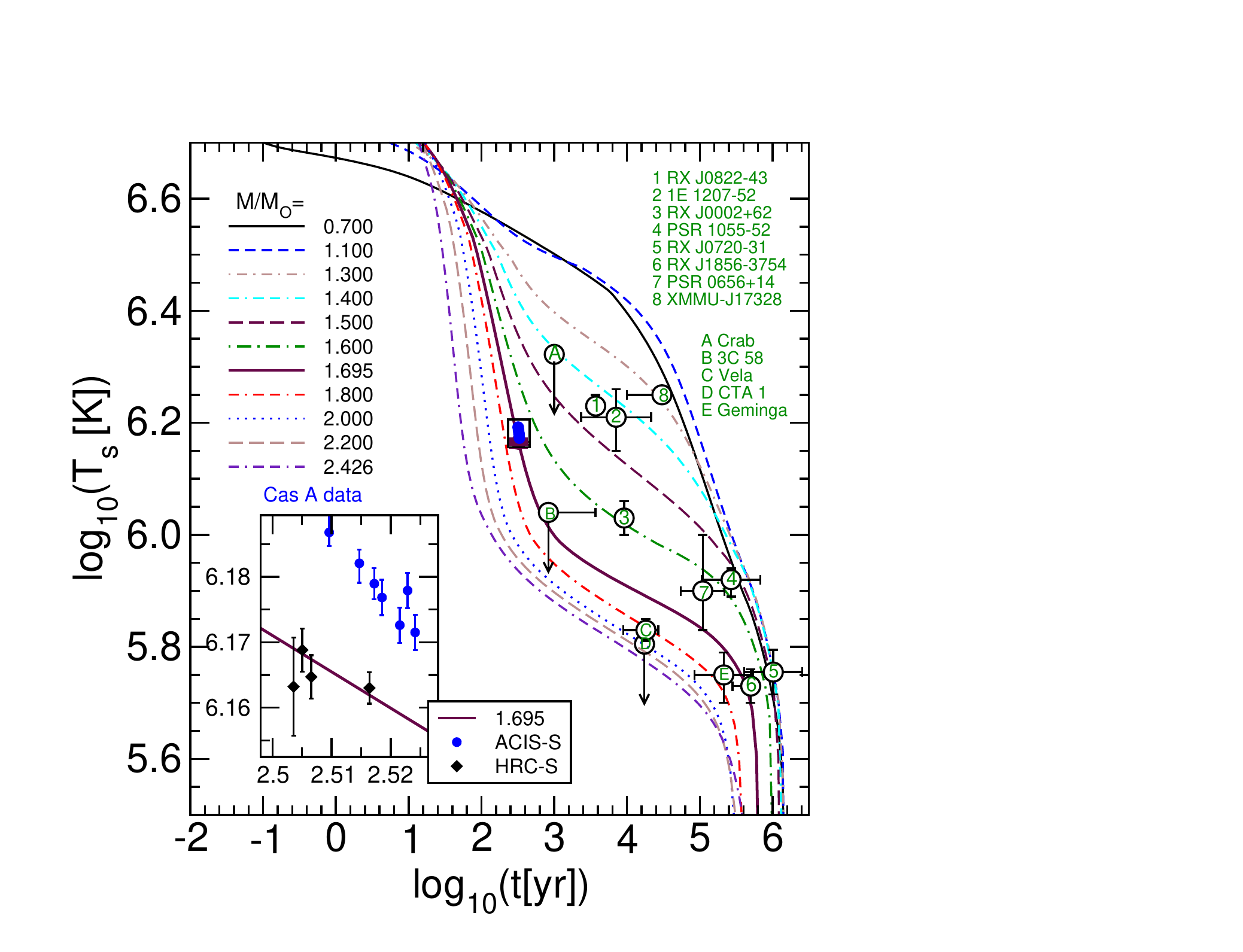}
      \caption{Same as Fig.~\ref{Fig:Cool2} for $n_c^{\pi}=2.0~n_0$.
Cas A cooling data from the HRC-S instrument are explained with a NS of $M=1.695~M_\odot$.}
   \label{Fig:Cool3a}
 \end{figure}

 \begin{figure}[!htb]
   \centering
   \includegraphics[width=0.75\textwidth,angle=0]{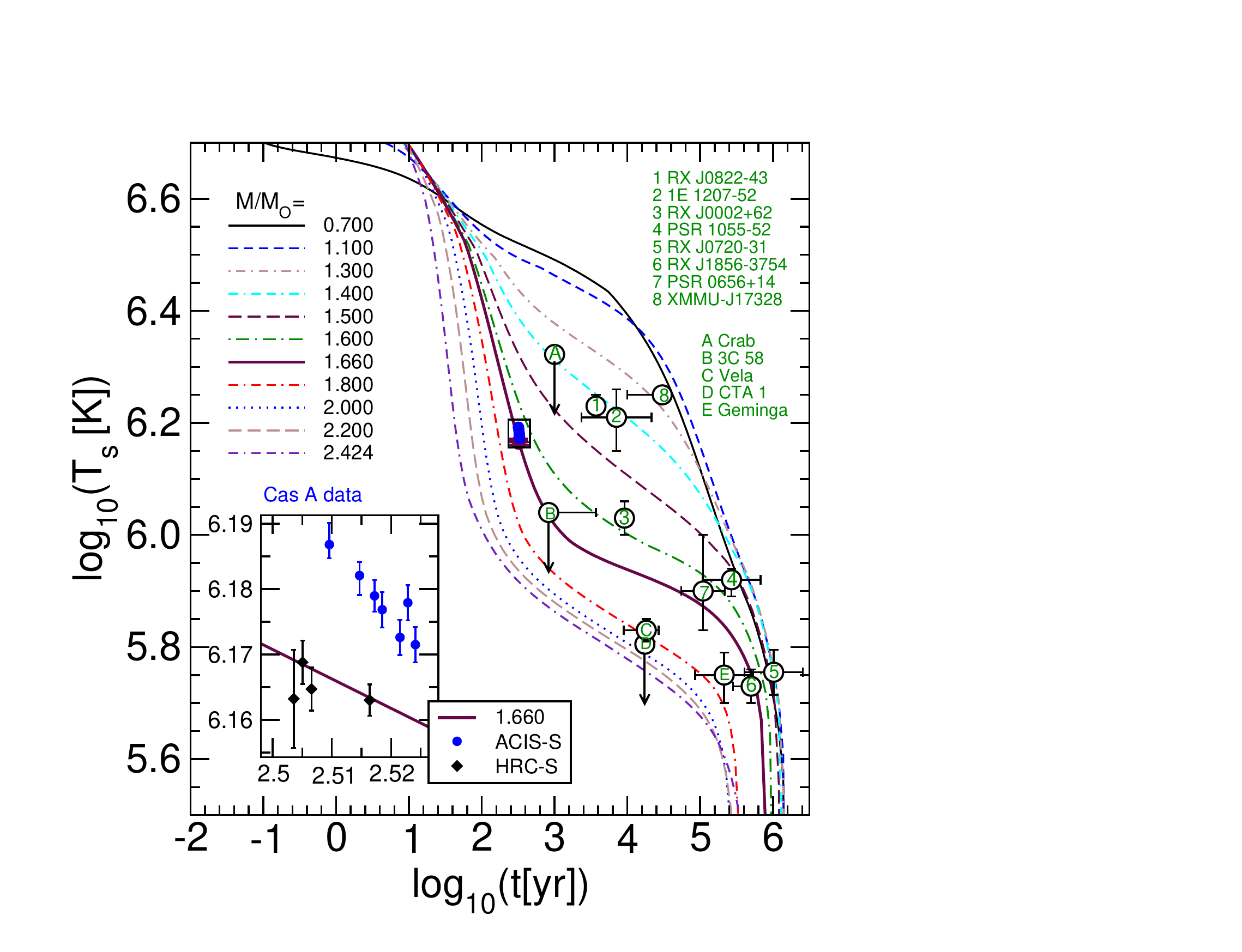}
      \caption{Same as Fig.~\ref{Fig:Cool2} for $n_c^{\pi}=1.5~n_0$.
      Cooling data for Cas A  from the HRC-S instrument are explained with a NS of $M=1.660~M_\odot$. }
   \label{Fig:Cool3b}
 \end{figure}

Now, exploiting the DD2 EoS we choose the same heat conductivity as in \cite{Blaschke:2013vma}
(without any additional tuning, i.e.  the lepton heat conductivity is computed
following \cite{Shternin:2007ee} and the nucleon contribution incorporates a
decrease with increasing density owing to the pion softening effect) 
and we tune the effective pion gap and the $pp$ pairing gap. 
This allows to describe the Cas~A cooling data by a NS with a lower mass. 
We performed calculations with all the $pp$ gap curves shown in
Fig.~\ref{Fig:gaps} and with $\omega^*(n)$ given by the various curves 1a+1b (i.e. suppressing the possibility of pion condensation) shown in Fig.~\ref{Fig:omegatilde}.  
To reproduce the cooling of Cas~A with a decline steeper than $2\%$ while simultaneously fitting other cooling data proves to be possible only for a couple of $pp$ gap and $\omega^*(n)$ choices. 
It is much easier to fit the set of cooling data when the decline of the Cas~A cooling data is less steep.

With the proton gap following models I and II, which we have used in case of the HDD EoS,
it is not possible to appropriately  fit the data using the DD2 EoS: the cooling proves to be too slow. 
One needs the $pp$ gaps dropping at a smaller density than in cases of the gaps given by models I and II.
Below we demonstrate cooling histories computed with EEHOr and  CCYms   $pp$ gaps,
which allow in two cases to get steep declines for Cas~A and with the BCLL gap, for which the above mentioned  decline is smaller. 
For the CCYms model the amplitude of the gap is smaller than that for the EEHOr model, but both gaps drop to zero at approximately the same density, $n\simeq 2.5~n_0$. 
The BCLL gap is less in amplitude than the EEHOr gap for $n>0.5~n_0$ and less than the CCYms gap 
for $n>0.8~n_0$; the BCLL gap drops to zero for $n\simeq 2.3~n_0$.

The resulting cooling curves are shown in Fig.~\ref{Fig:Cool2}
for the $pp$ pairing gap of the model EEHOr from Fig.~\ref{Fig:gaps} and for the effective pion gap given by the solid curve 1a+1b, $n_c^{\pi}=3~n_0$, from Fig.~\ref{Fig:omegatilde}. 
At rather low densities relevant for stars with a mass $M\lsim M_{\odot}$
the pion softening effect is not pronounced. 
In the EEHOr model the $pp$ proton gap is  large at such densities.  
Thus the mass of the hot objects proves to be essentially above $M_{\odot}$.
The description of the coldest among not very old objects ``C and D'' requires a $pp$ gap dropping to zero at central densities reached in those massive objects. 
The cooling of these objects is determined by the efficient MMU process in the absence of pairing. 
The decline of the curve describing Cas~A  proves to be  $1.9\%$.  
Cas~A cooling data from the ACIS-S instrument are explained with a NS of $M=1.950~M_\odot$ while
data from the HRC-S instrument are explained with a slightly more massive star of $M=1.980~M_\odot$. The whole range of data is covered by the cooling curves in the mass interval from $\sim 1.4$ to 
$2.43~M_{\odot}$.  
The Crab pulsar is described by $\geq 1.6~M_{\odot}$ and Vela by $2.2 M_{\odot}$. 
The XMMU source has a mass of $1.42-1.55~M_{\odot}$. 
All other sources, except  "C" and "D''  are covered by stars with masses from 1.4 to 1.95 $M_{\odot}$. 
Thus, with the given model  the XMMU, Crab, Cas A and Vela objects prove to be very  massive.
 \begin{figure}[!htb]
   \centering
   \includegraphics[width=0.75\textwidth,angle=0]{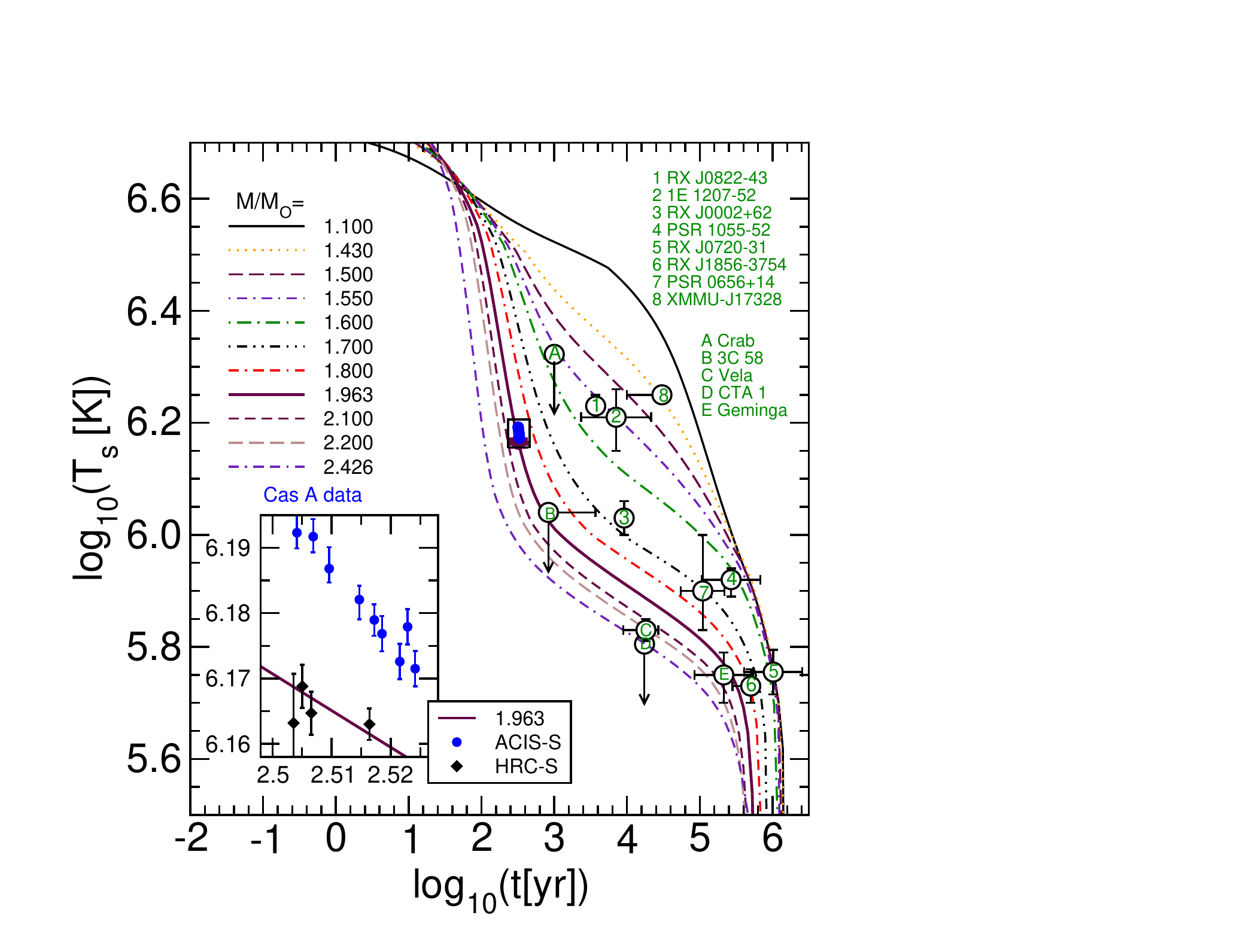}
      \caption{Same as Fig.~\ref{Fig:Cool2}, but for the $1S_0$ $pp$ pairing gap model CCYms.  
Cas~A cooling data from the HRC-S instrument are explained with a NS of $M=1.963~M_\odot$. }
   \label{Fig:Cool4}
 \end{figure}

  \begin{figure}[!htb]
   \centering
   \includegraphics[width=0.75\textwidth,angle=0]{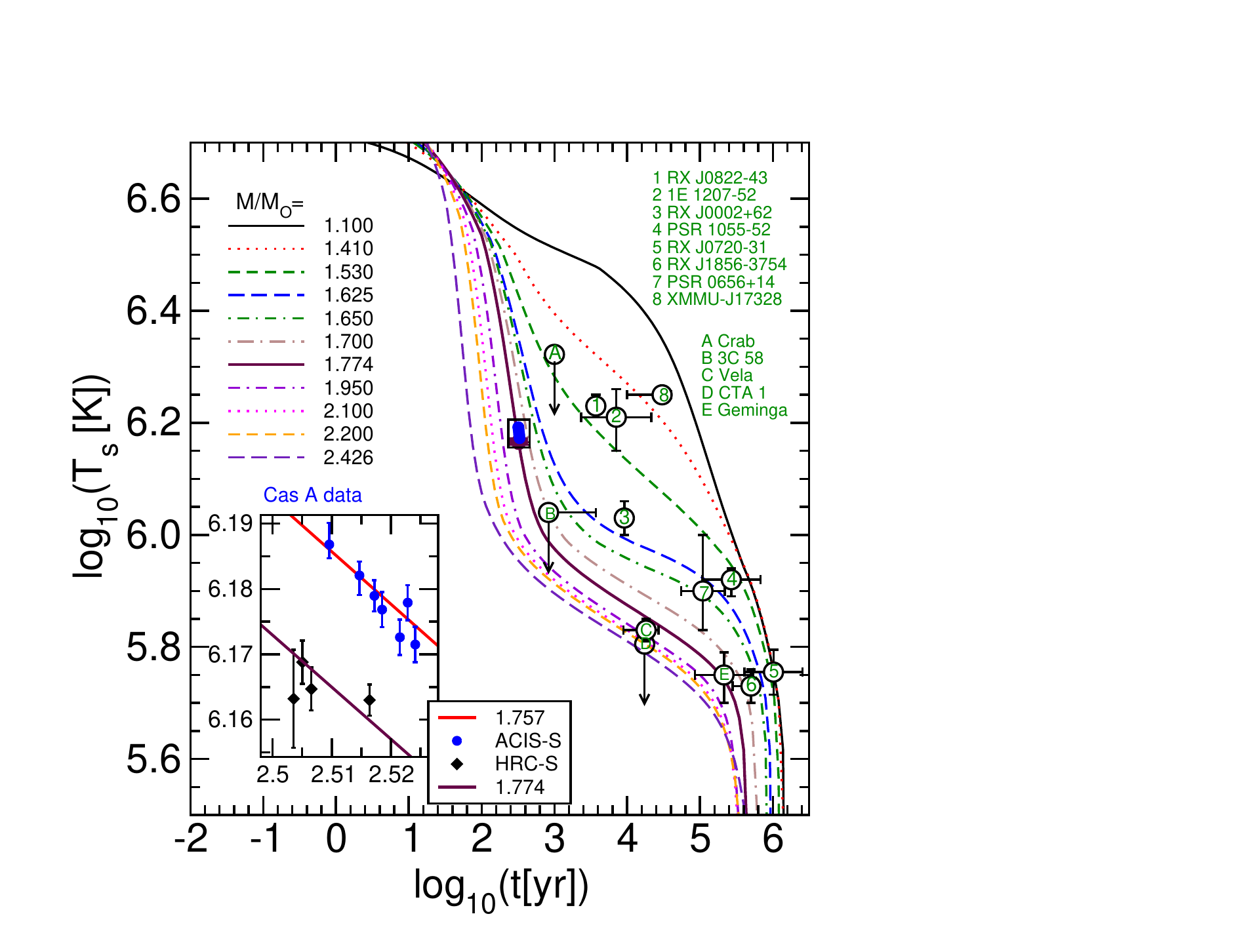}
      \caption{Same as Fig.~\ref{Fig:Cool4}, but for $n_c^{\pi}=2.5~n_0$.
Cas~A cooling data from the HRC-S instrument are
explained with a NS of $M=1.774~M_{\odot}$, and data from ACIS-S with $M=1.757~M_{\odot}$. }
   \label{Fig:Cool5}
 \end{figure}
   \begin{figure}[!htb]
   \centering
   \includegraphics[width=0.75\textwidth,angle=0]{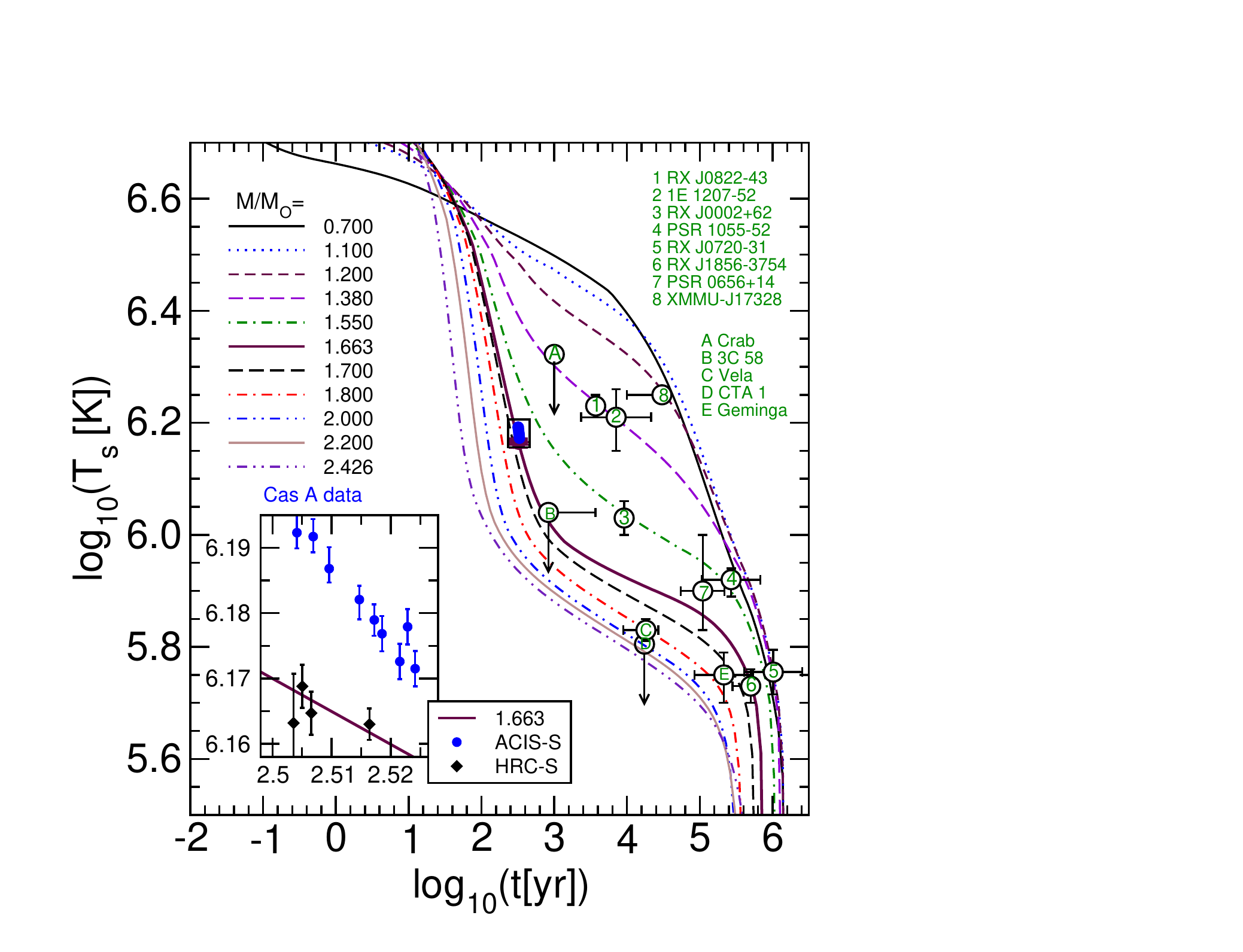}
      \caption{Same as Fig.~\ref{Fig:Cool4}, but for $n_c^{\pi}=2.0~n_0$.
Cas~A cooling data from the HRC-S instrument are
explained with NS of $M=1.663~M_\odot$. }
   \label{Fig:Cool5a}
 \end{figure}
    \begin{figure}[!htb]
   \centering
   \includegraphics[width=0.75\textwidth,angle=0]{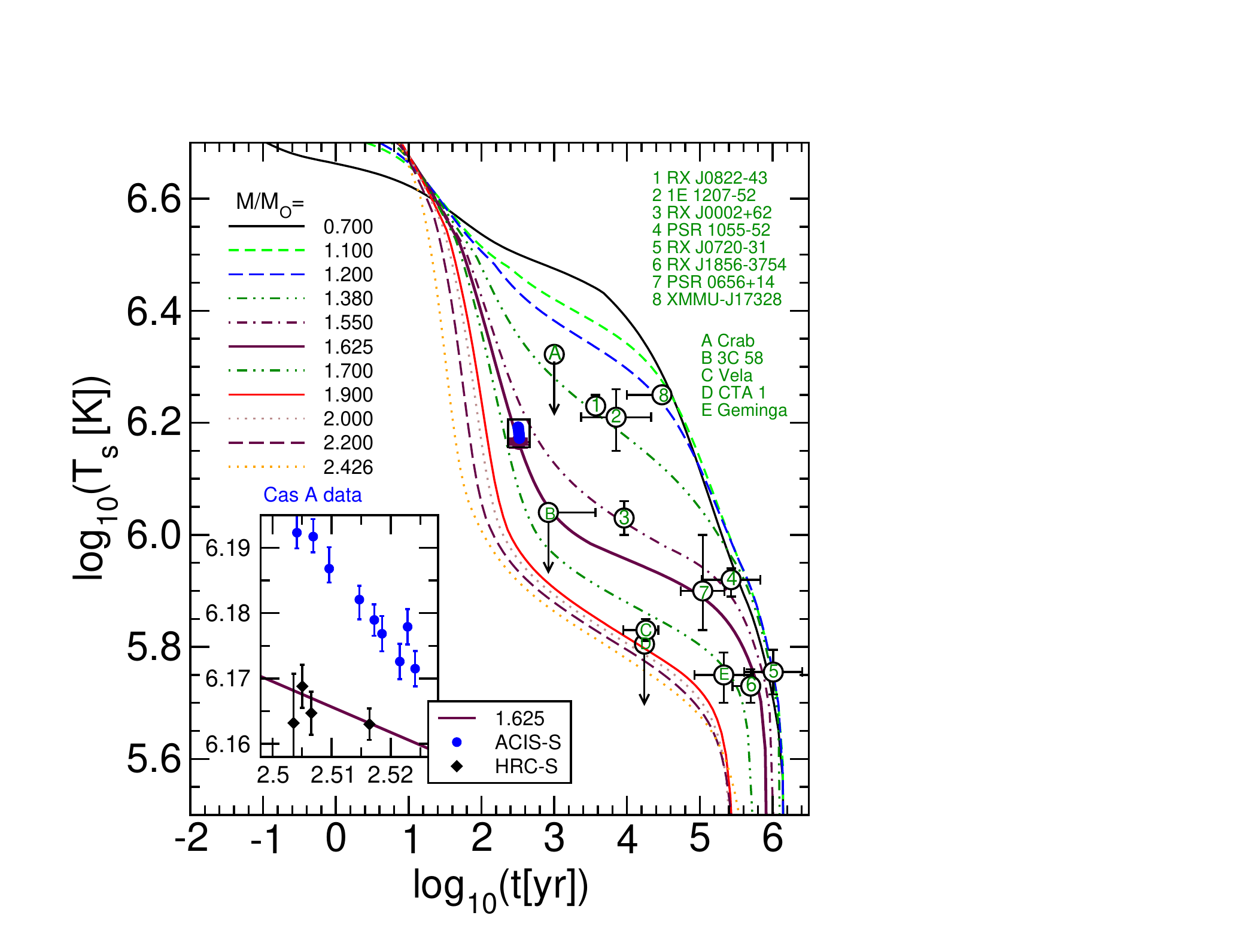}
      \caption{Same as Fig.~\ref{Fig:Cool4}, but for $n_c^{\pi}=1.5~n_0$.
Cas~A cooling data from the HRC-S instrument are
explained with NS of $M=1.625~M_\odot$. }
   \label{Fig:Cool5b}
 \end{figure}

In order to describe the cooling data from Cas~A with a less massive neutron star, a smaller value of the effective pion gap can be used. 
In Fig.~\ref{Fig:Cool3} we show the same as in Fig.~\ref{Fig:Cool2} but with the effective pion gap calculated with the help of the dashed curve 1a+1b in Fig.~\ref{Fig:omegatilde}, for $n_c^{\pi}=2.5~n_0$. The cooling of the hot source XMMU is now explained by a NS with a  mass of $1.4-1.47~ M_{\odot}$, and the whole range of data is covered by the cooling curves in the mass interval from $\sim$ 1.4 to $2.43 M_{\odot}$. 
We see that the Cas~A cooling data can now be explained with a lower NS mass of  $M=1.780~M_\odot$.  The decline of the curve describing Cas~A proves to be $2.6\%$ over 10 years in this model, being compatible with ACIS-S data. 
The Crab pulsar is described by a mass exceeding $1.5~M_{\odot}$ and Vela by $M\sim 1.9~M_{\odot}$.
All  sources, except  "C" and "D''  are covered by stars with masses from 1.4 to 1.8 $M_{\odot}$. 
So, if the Cas~A data supported a steep decline of the  cooling curve and the NS radii were $\gsim 13$ km, it could be considered as an argument in favour of  this model.

In Fig.~\ref{Fig:Cool3a} we show the same as in Fig.~\ref{Fig:Cool2} and in Fig.~\ref{Fig:Cool3}, but now the effective pion gap is described by the dotted curve 1a+1b in Fig.~\ref{Fig:omegatilde}, for 
$n_c^{\pi}=2~n_0$. 
The cooling of the hot source XMMU is  explained by a NS with a  mass of $1.3-1.35~M_{\odot}$, and the whole range of data is covered by cooling curves for stars in the mass interval from $\sim$ 1.3 to 
$2.43~M_{\odot}$. 
We see that the Cas~A cooling data can now be explained with a still lower NS mass of  
$M=1.695~M_\odot$.  
The decline rate of the curve describing Cas~A HRC-S data proves to be $1.7\%$ per decade in this model. 
The Crab pulsar is described by $M>1.4 M_{\odot}$ and Vela by $M\sim 1.8 M_{\odot}$. 
All  sources, except the most rapidly cooling objects "C" and "D'',  are covered by stars with masses from 
1.3 to 1.75 $M_{\odot}$.

Fig.~\ref{Fig:Cool3b} shows the same as  Fig.~\ref{Fig:Cool2}, Fig.~\ref{Fig:Cool3} and Fig.~\ref{Fig:Cool3a} but with the effective pion gap given by the dash-dotted curve 1a+1b in Fig.~\ref{Fig:omegatilde}, with $n_c^{\pi}=1.5 n_0$, corresponding to the most efficient pion softening effect. 
The cooling of the hot source XMMU is now explained by a NS with a  mass of $1.2-1.35~M_{\odot}$, 
and the whole range of data is covered by the cooling curves for NS in the mass interval from $\sim$ 1.2 to $2.43~M_{\odot}$.
We see that the Cas A cooling data can now be explained with a lower NS mass of  $M=1.660~M_\odot$.  The decline of the curve describing Cas~A proves to be $1.4\%$ per decade. 
The Crab pulsar is described by $M>1.4 M_{\odot}$ and Vela by $M\sim 1.8~M_{\odot}$. 
All  sources, except  most rapidly cooling objects "C" and "D'',  are covered by stars with masses from 1.2 to 1.7 $M_{\odot}$.

Thus with the model EEHOr for the $pp$ gaps we are able to describe the whole set of the data and could chose between different effective pion gap curves if the decline of the curve describing Cas~A were known. From the point of view of a similarity of the distribution of the sources to the one that follows from a population synthesis \cite{Popov:2004ey} and supernova simulation models \cite{Ugliano:2012kq},
the models shown in Figs.~\ref{Fig:Cool3a} and \ref{Fig:Cool3b} are more appropriate than those shown in Figs.~\ref{Fig:Cool2} and \ref{Fig:Cool3}. 
Therefore, an appropriate description of NS cooling with stiff EoS can be considered as an argument for a  strong pion softening.

\begin{figure}[!htb]
   \centering
   \includegraphics[width=0.75\textwidth,angle=0]{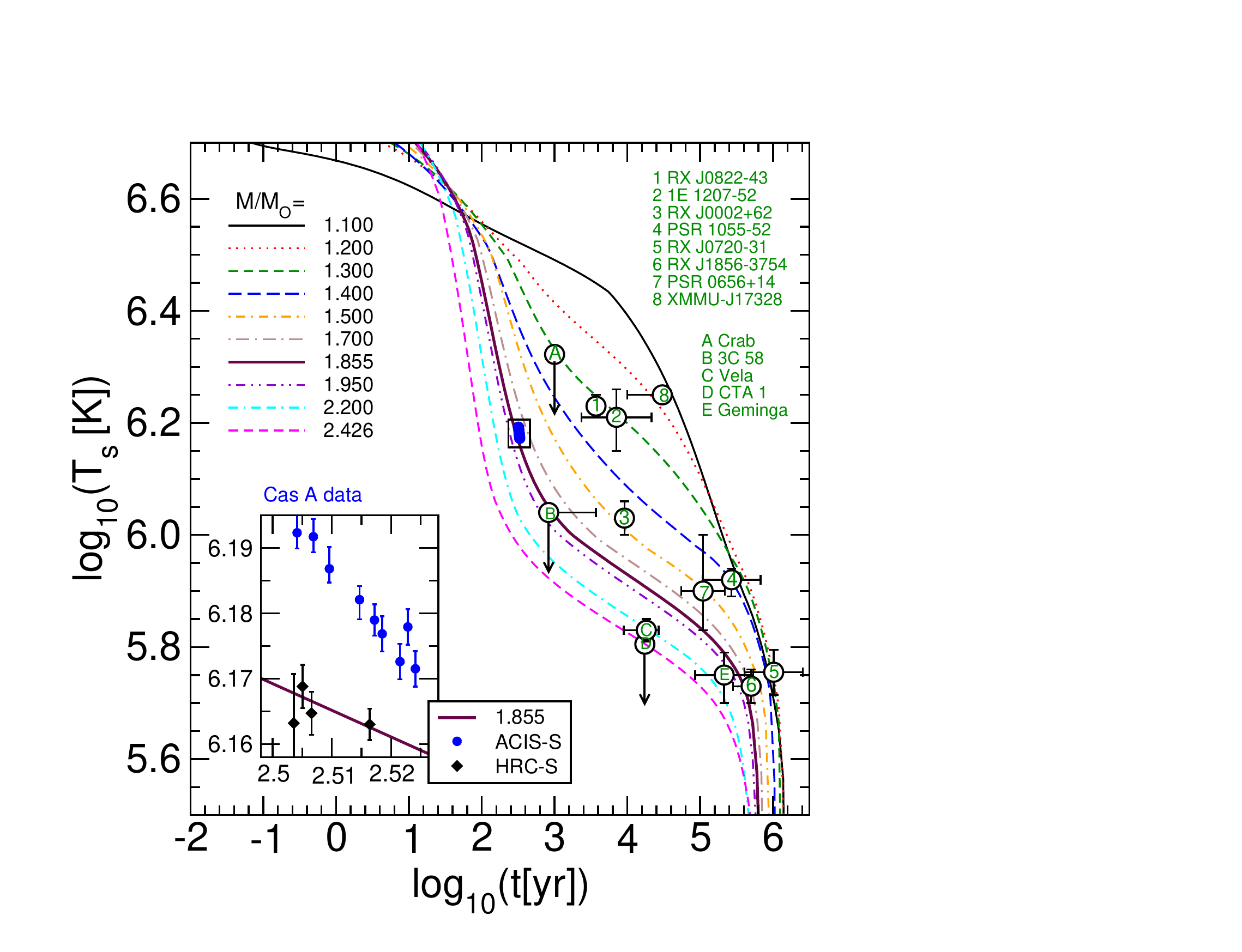}
      \caption{Same as Fig.~\ref{Fig:Cool4}, but for the $1S_0$ $pp$ pairing gap model BCLL.
      Cas~A cooling data from the HRC-S instrument are explained with a NS of $M=1.855~M_\odot$.   }
   \label{Fig:Cool6}
 \end{figure}

\begin{figure}[!htb]
   \centering
   \includegraphics[width=0.75\textwidth,angle=0]{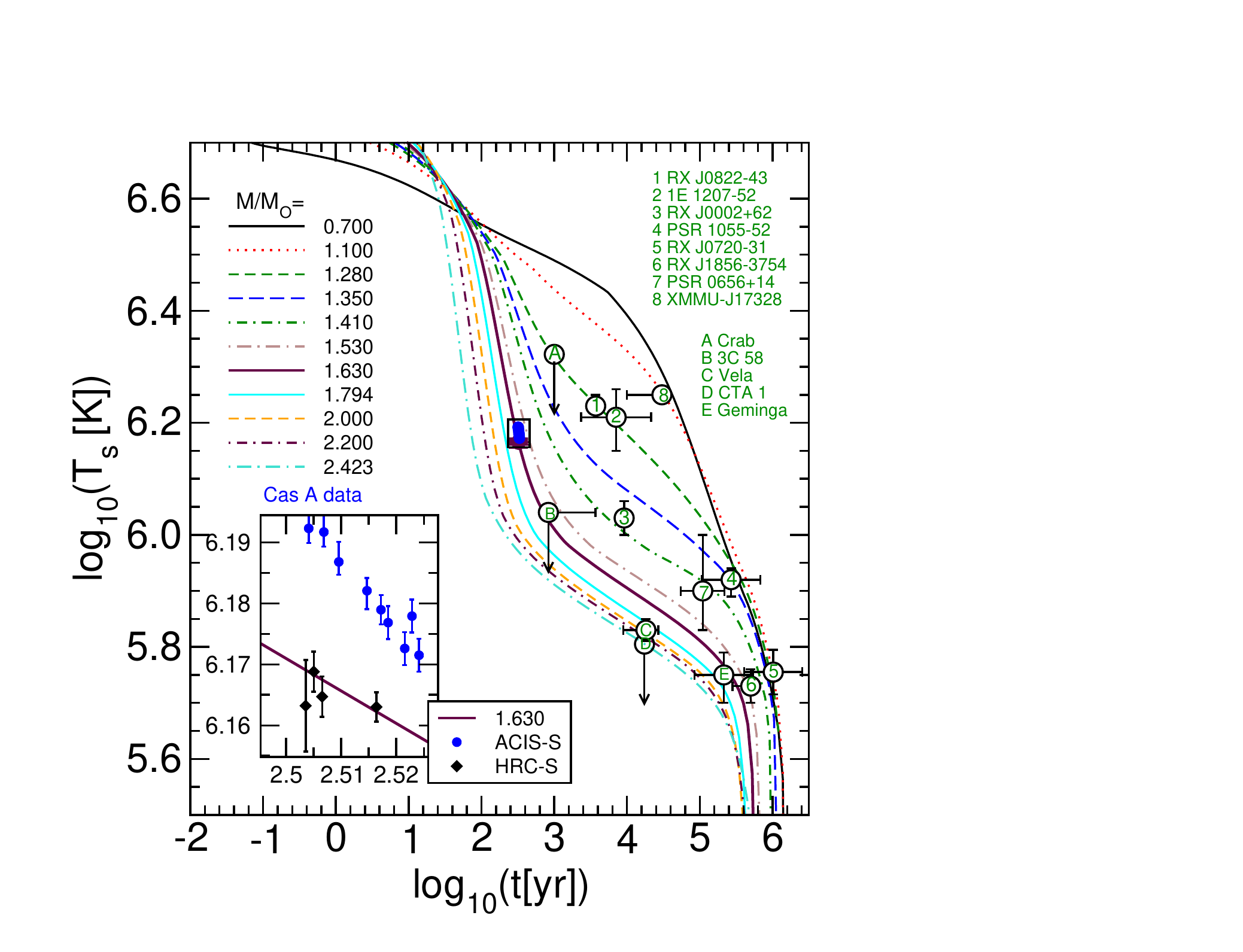}
      \caption{Same as Fig.~\ref{Fig:Cool6}, but for $n_c^{\pi}=2.5~n_0$. 
      Cas~A cooling data from the HRC-S instrument are 
explained with a NS of $M=1.630~M_\odot$.   }
   \label{Fig:Cool7}
 \end{figure}
\begin{figure}[!htb]
   \centering
   \includegraphics[width=0.75\textwidth,angle=0]{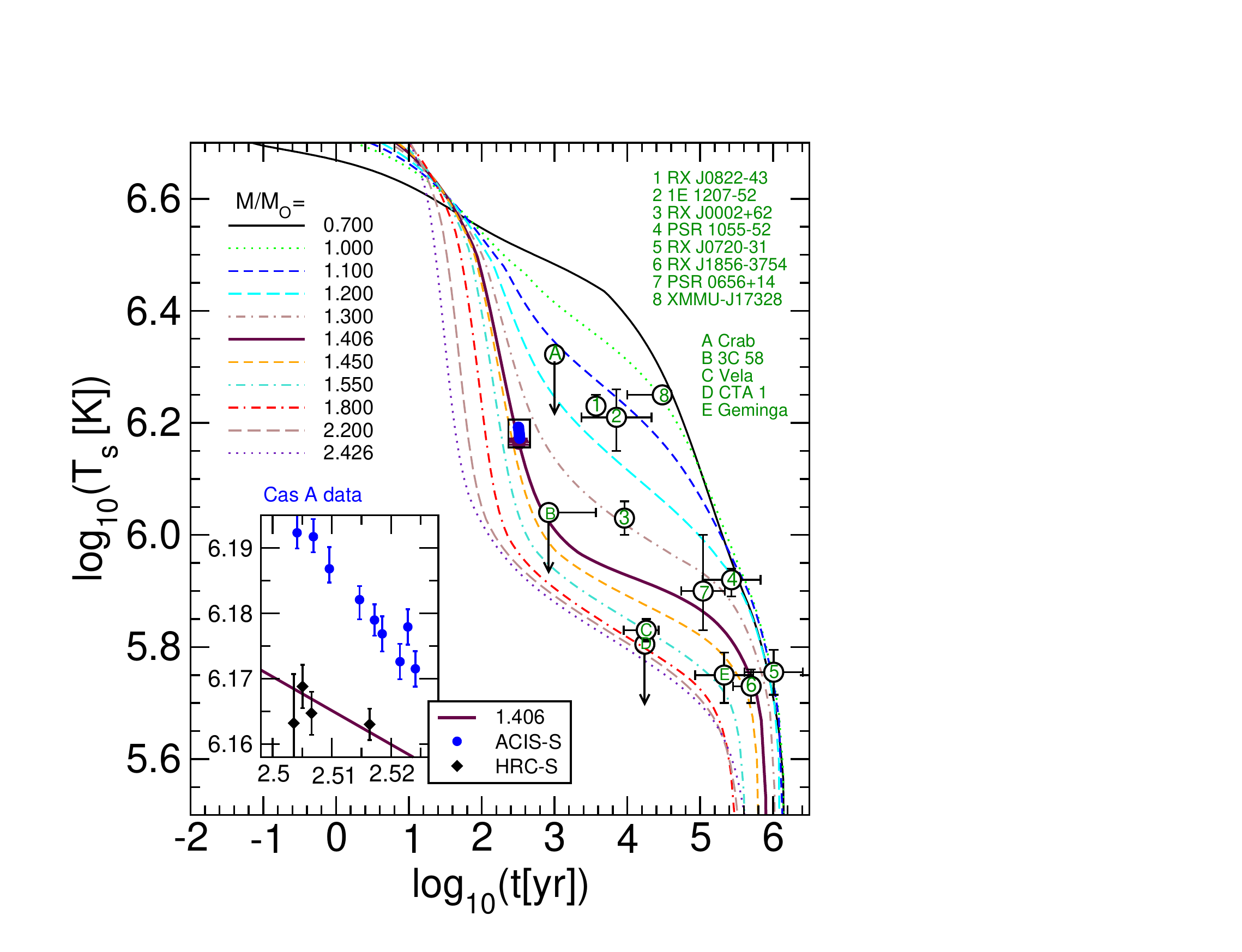}
      \caption{Same as Fig.~\ref{Fig:Cool6}, but for $n_c^{\pi}=2.0~n_0$. 
      Cas~A cooling data from the HRC-S instrument are 
explained with a NS of $M=1.406~M_\odot$.   }
   \label{Fig:Cool7a}
 \end{figure}
 \begin{figure}[!htb]
   \centering
   \includegraphics[width=0.75\textwidth,angle=0]{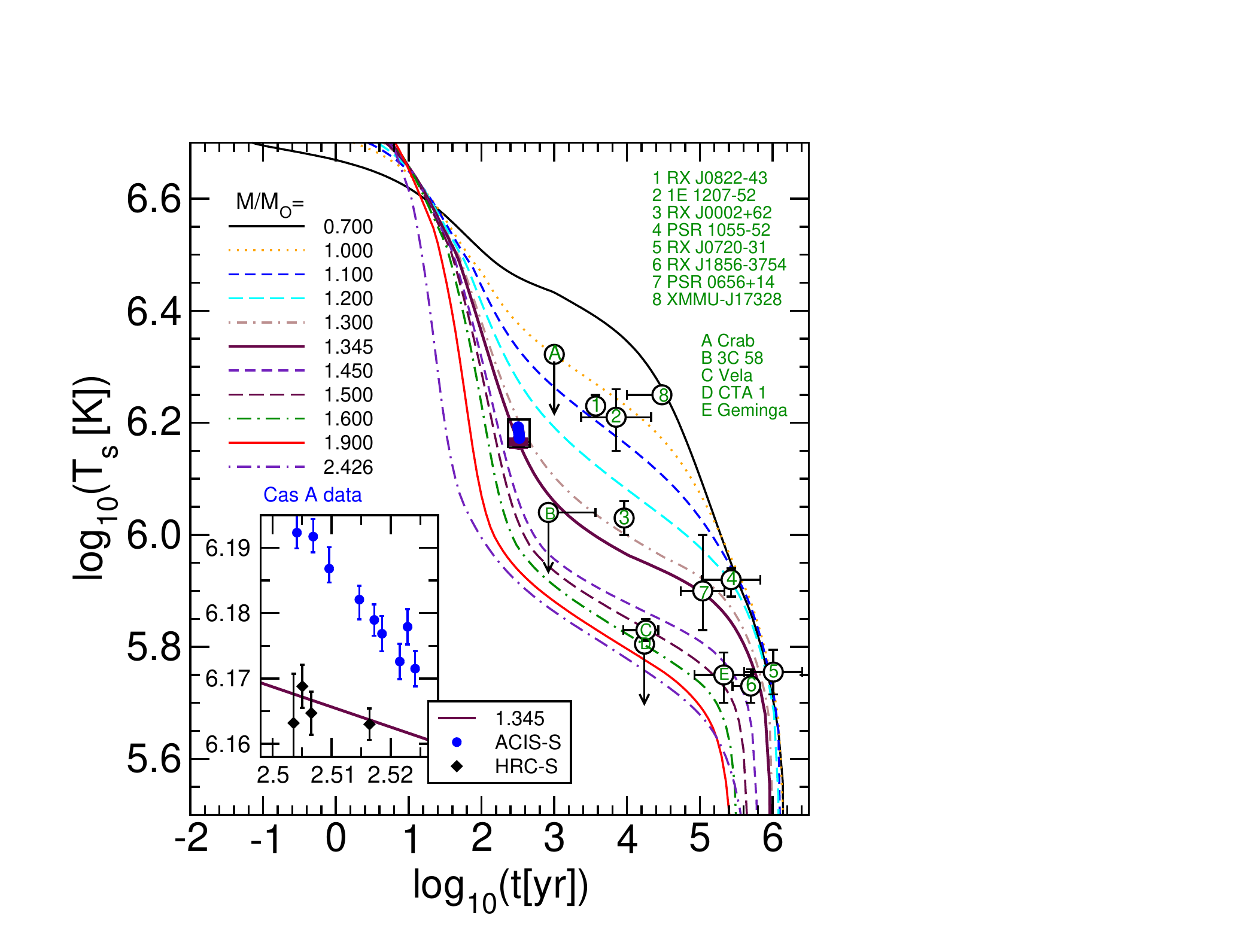}
      \caption{Same as Fig.~\ref{Fig:Cool6}, but for $n_c^{\pi}=1.5~n_0$. 
      Cas~A cooling data from the HRC-S instrument are 
explained with a NS of $M=1.345~M_\odot$.   }
   \label{Fig:Cool7b}
 \end{figure}
 \begin{figure}[!htb]
   \centering
   \includegraphics[width=0.75\textwidth,angle=0]{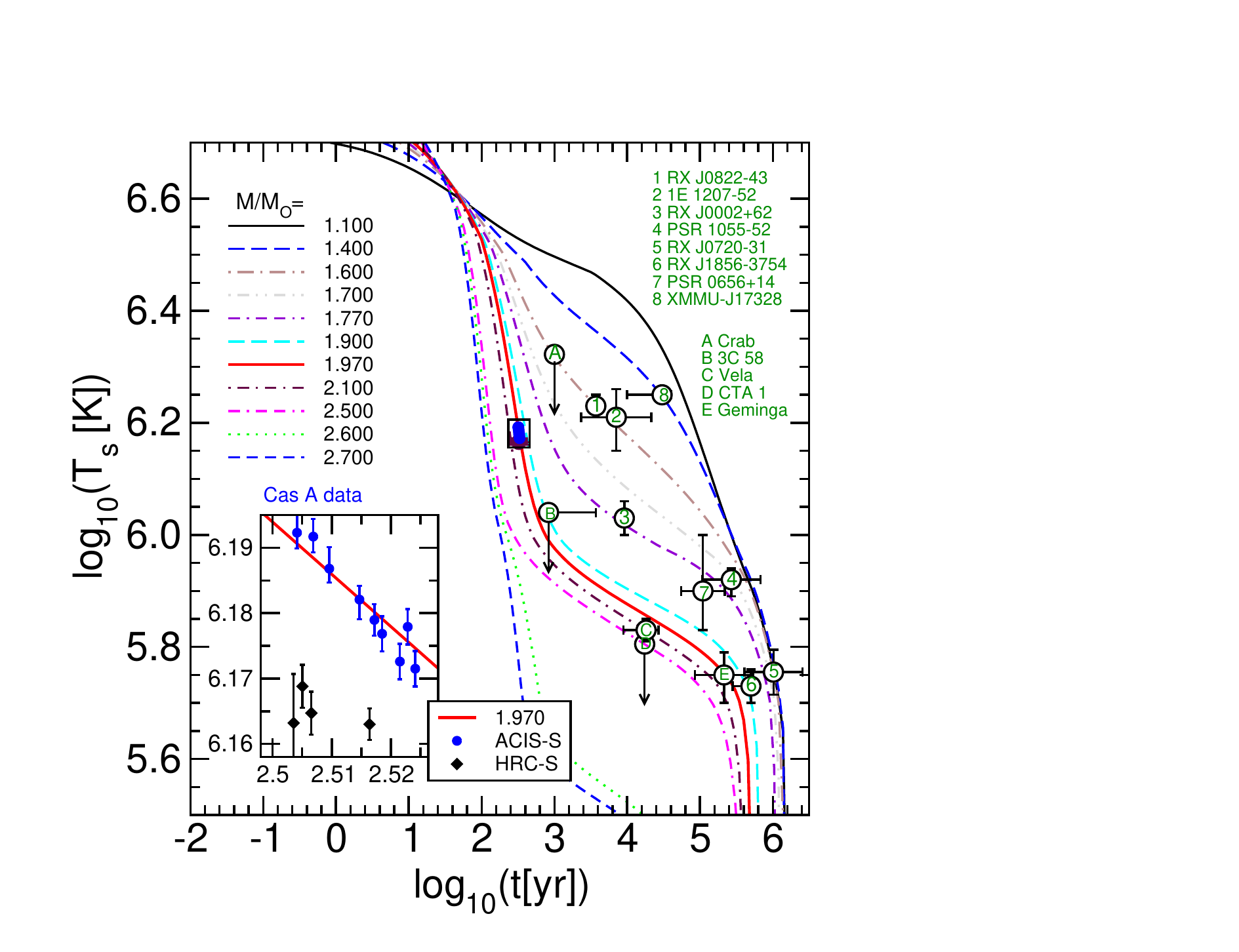}
      \caption{Same as Fig.~\ref{Fig:Cool7a}, but for the  hadronic
      DD2vex  EoS.
      Cas~A cooling data from the ACIS-S instrument in graded mode are 
explained with a NS of $M=1.970~M_\odot$.   }
   \label{Fig:Cool8}
 \end{figure}
 \begin{figure}[!htb]
   \centering
   \includegraphics[width=0.75\textwidth,angle=0]{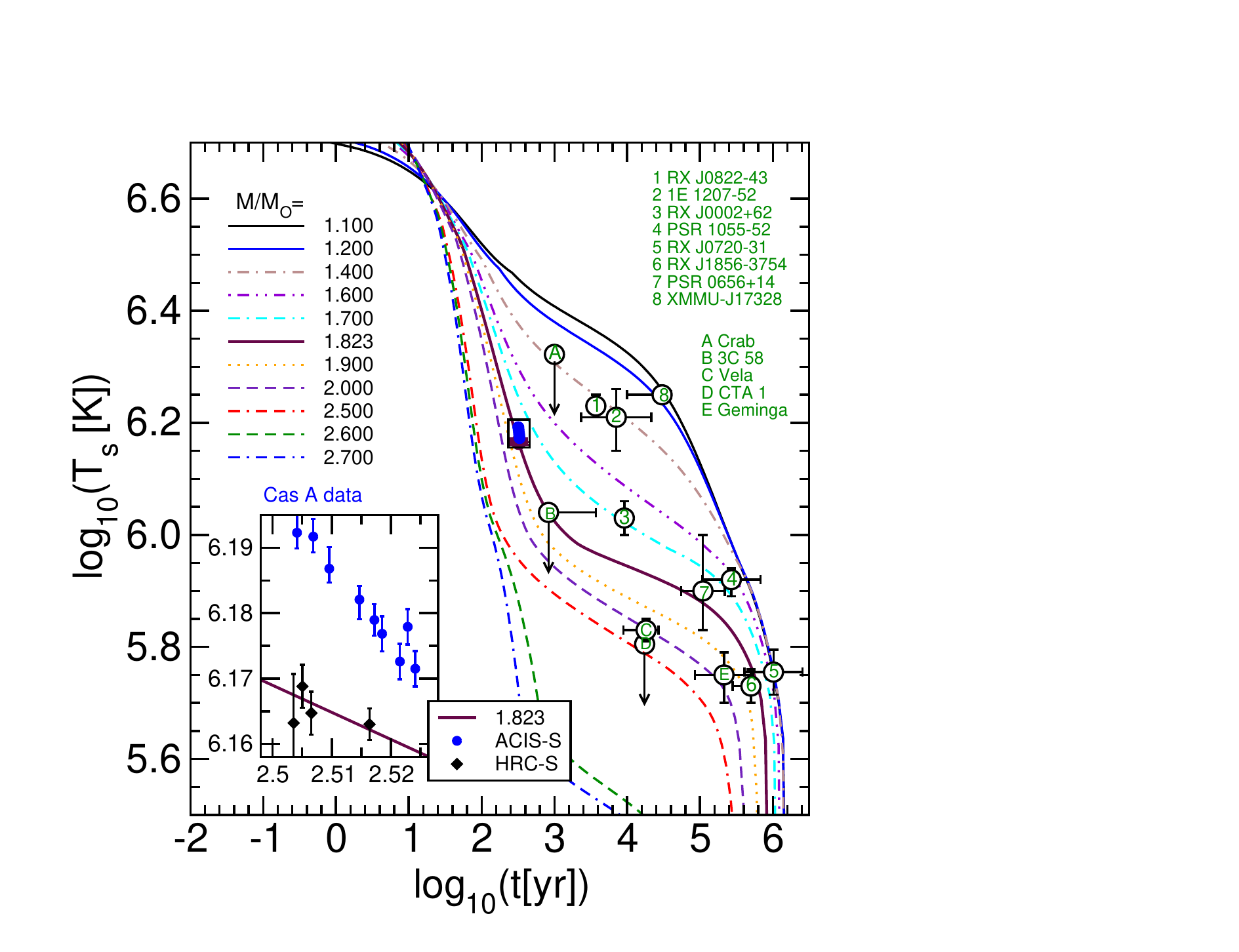}
      \caption{Same as Fig.~\ref{Fig:Cool8}, but for $n_c^{\pi}=1.5~n_0$.
      Cas~A cooling data from the HRC-S instrument are 
explained with a NS of $M=1.823~M_\odot$.   }
   \label{Fig:Cool8a}
 \end{figure}
   \begin{figure}[!htb]
   \centering
   \includegraphics[width=0.75\textwidth,angle=0]{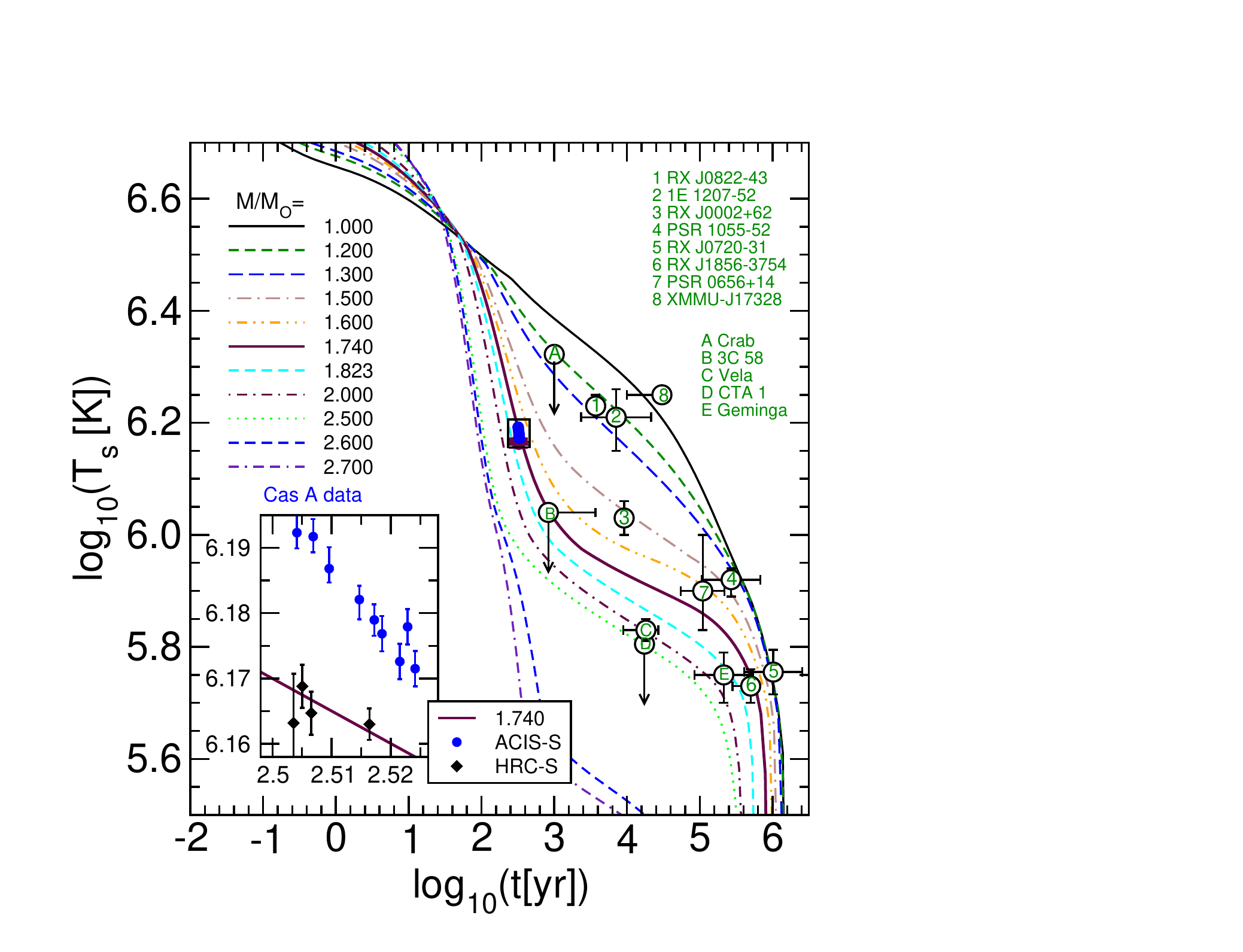}
      \caption{Same as Fig.~\ref{Fig:Cool8}, but for The $1S_0$ $pp$ pairing gap model AO. 
      Cas~A cooling data from the HRC-S instrument are 
explained with a NS of $M=1.740~M_\odot$.   }
   \label{Fig:Cool8b}
 \end{figure}
  \begin{figure}[!htb]
   \centering
   \includegraphics[width=0.75\textwidth,angle=0]{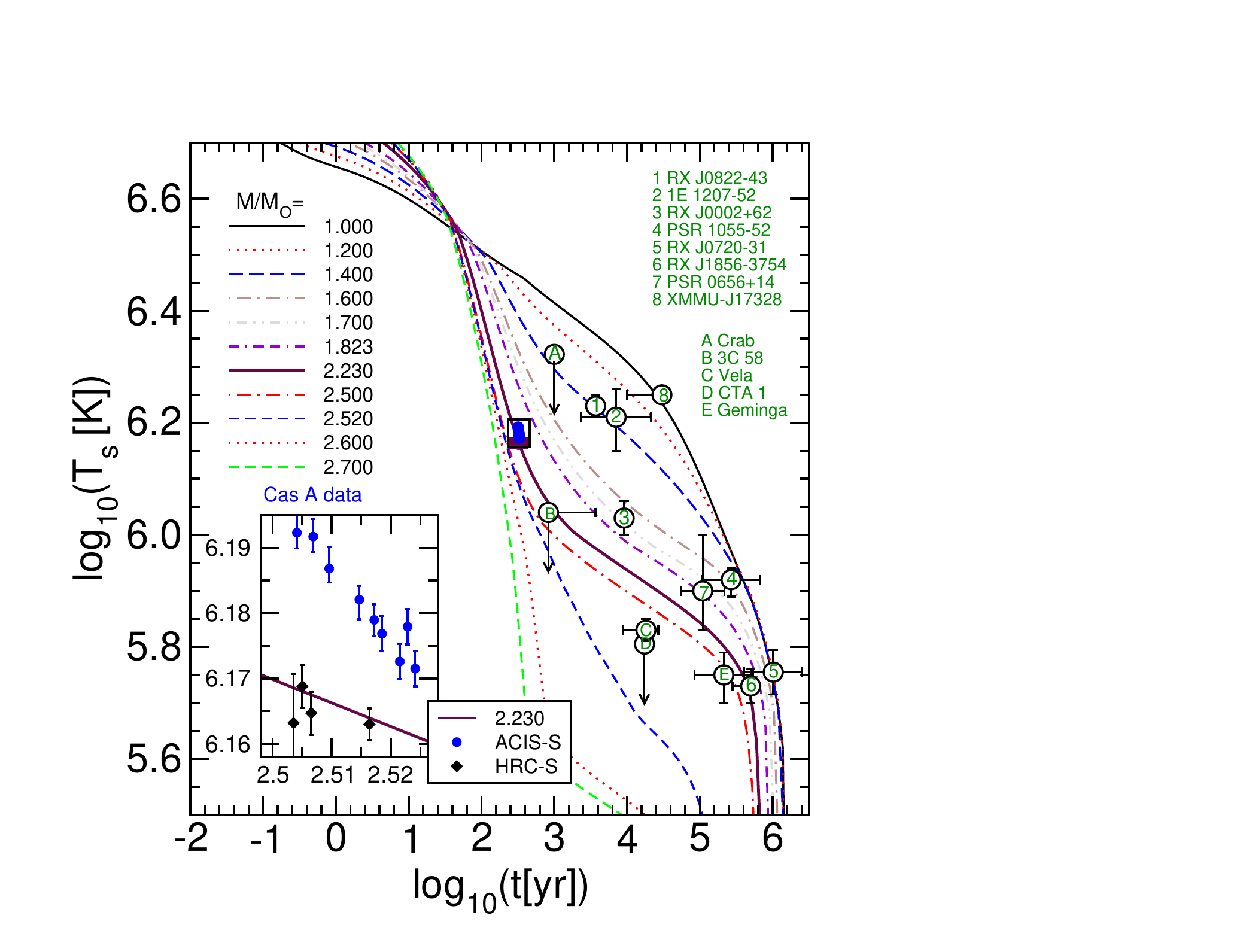}
      \caption{Same as Fig.~\ref{Fig:Cool8b}, but for $n_c^{\pi}=2.5~n_0$. 
      Cas~A cooling data from the HRC-S instrument are 
explained with a NS of $M=2.230~M_\odot$.   }
   \label{Fig:Cool8c}
 \end{figure}
In Fig.~\ref{Fig:Cool4} we show the same as in Fig. \ref{Fig:Cool2} ($n_c^{\pi}= 3~n_0$)  but for the  $1S_0$ $pp$ pairing gap corresponding to the model CCYms. 
With these $pp$ pairing gaps we reproduce Cas~A cooling data from the HRC-S instrument with a NS mass of $M=1.963~M_\odot$, which is only slightly different from $M=1.980~M_\odot$, with which we  reproduced the HRC-S data in the example shown in Fig.~\ref{Fig:Cool2} for the EEHOr $pp$ gap model.  The decline of the curve describing Cas~A data is $1.8\%$ per decade. 
The whole range of data is covered by cooling curves in the mass interval from $\sim 1.4$ to $2.43 M_{\odot}$. 
The Crab pulsar is described by $M>1.55~M_{\odot}$ and Vela by $\sim 2.2~M_{\odot}$. 
The XMMU source has a mass of $1.4-1.5~M_{\odot}$. 
All  sources, except "C" and "D''  are covered by stars with masses from 1.4 to 1.96 $M_{\odot}$.
So, the model has the same deficiencies as the model EEHOr for $n_c^{\pi}=3~n_0$: 
the hottest objects and intermediate cooling objects are very massive.

In Fig.~\ref{Fig:Cool5} we show the same as in Fig.~\ref{Fig:Cool4}, but for $n_c^{\pi}= 2.5~n_0$. 
We reproduce the Cas~A cooling data from the HRC-S instrument for a NS mass $M=1.774~M_\odot$ and ACIS-S data for a NS mass $M=1.757~M_\odot$, which is only slightly different from $M=1.780~M_\odot$, with which we  reproduced ACIS-S data in Fig.~\ref{Fig:Cool3} with EEHOr gaps.  
The decline of the curve describing Cas~A decreased up to $2.3\%$ per decade. 
The whole range of data is covered by cooling curves in the mass interval from $\sim 1.3$ to 
$2.43~M_{\odot}$. 
The Crab pulsar is described by $M>1.5 M_{\odot}$ and Vela by $M\sim 1.95~M_{\odot}$. 
The XMMU source has a mass $1.3-1.45~M_{\odot}$. 
All  sources, except the most rapidly cooling objects "C" and "D''  are covered by stars with masses from 
1.3 to 1.7 $M_{\odot}$.

In Fig.~\ref{Fig:Cool5a} we show the same as in Fig. \ref{Fig:Cool4} but for $n_c^{\pi}=2~n_0$. 
Now we reproduce HRC-S data for Cas~A with a NS mass of $M=1.663~M_\odot$ rather than $M=1.695~M_\odot$ as in Fig.~\ref{Fig:Cool3a}.  
The decline of the curve describing Cas~A is $1.7\%$ per decade, as it was in  case shown in  
Fig.~\ref{Fig:Cool3a} for the EEHOr $pp$ gap. 
The whole range of data is covered by cooling curves in the mass interval from $\sim 1.2$ to 
$2.43~M_{\odot}$. 
The Crab pulsar is described by $M>1.35 M_{\odot}$ and Vela by $M\sim 1.8~M_{\odot}$. 
The XMMU source has a mass of $1.2-1.25~M_{\odot}$. 
All  sources, except the most rapidly cooling objects  "C" and "D''  are covered by stars with masses 
from 1.2 to 1.7 $M_{\odot}$.

In Fig.~\ref{Fig:Cool5b} we show the same as in Fig.~\ref{Fig:Cool4} but for $n_c^{\pi}=1.5~n_0$ 
With the $pp$ pairing gaps corresponding to model CCYms we reproduce the HRC-S data  for a NS mass $M=1.625~M_\odot$ rather than $M=1.660~M_\odot$ as in  Fig.~\ref{Fig:Cool3b} for the $pp$ pairing gaps corresponding to model EEHOr.  
The decline of the curve describing Cas~A is $1.3\%$ per 10 years. 
The whole range of data is covered by cooling curves in the mass interval from $\sim 1.1$ to 
$2.43~M_{\odot}$. 
The Crab pulsar is described by $M>1.3 M_{\odot}$ and Vela by $M\sim 1.8 M_{\odot}$. 
The XMMU source has a mass of $1.1-1.25~M_{\odot}$. 
All  sources, except the most rapidly cooling objects "C" and "D'',  are covered by stars with masses from 
1.1 to 1.7 $M_{\odot}$.

Thus we are able to conclude that descriptions of the data with the $pp$ gaps given by EEHOr and 
CCYms models are rather similar. 
Only the model $n_c^{\pi}=3n_0$ might be excluded in both cases, since all the objects prove to be too heavy.  
If the Cas~A data supported a steep decline of the  cooling curve and the NS radii were $\gsim 13$ km, 
it could be considered as an argument in favour of  the models presented in Figs. ~\ref{Fig:Cool2}, ~\ref{Fig:Cool3} and ~\ref{Fig:Cool5}.

In Fig.~\ref{Fig:Cool6} we show the same as in Fig. \ref{Fig:Cool2} and Fig.~\ref{Fig:Cool4} 
($n_c^{\pi}=3~n_0$),   but for the  $1S_0$ $pp$ pairing gaps corresponding to model BCLL. 
The BCLL gaps have approximately the same values as the CCYms ones but are shifted to smaller densities. 
With these gaps we reproduce the HRC-S Cas A data for a NS mass of $M=1.855~M_\odot$ and obtain a decline of $1.3\%$ per decade, being smaller than in the cases shown in Fig.~\ref{Fig:Cool2} and 
Fig.~\ref{Fig:Cool4}. 
The whole range of the data is covered by cooling curves in a broad interval of  masses from 1.1 to 
$2.43~M_{\odot}$. 
The XMMU source has a mass $1.1-1.25~M_{\odot}$. 
The Crab pulsar is described by $M>1.3 M_{\odot}$ and Vela by $M\sim 2.2 M_{\odot}$. 
All  sources, except the most rapidly cooling objects "C" and "D''  are covered by stars with masses from 
1.1 to 1.95 $M_{\odot}$.
So, the model has the same deficiencies as the  EEHOr and CCYms models for $n_c^{\pi}=3~n_0$.

In Fig.~\ref{Fig:Cool7} we show the same as in Fig.~\ref{Fig:Cool3} and Fig.~\ref{Fig:Cool5} 
($n_c^{\pi}=2.5~n_0$),   but for the  $1S_0$ $pp$ pairing gaps corresponding to model BCLL. 
Now we reproduce the HRC-S data for a NS mass $M=1.630~M_\odot$ and obtain a decline of $1.6\%$
per decade. 
Note that in the examples shown in Fig.~\ref{Fig:Cool3} and Fig.~\ref{Fig:Cool5} we were able to reproduce ACIS-S data with masses $M=1.780~M_\odot$ and $M=1.757~M_\odot$ respectively. 
The whole range of the data is covered by cooling curves for masses from 1.1 to $2.43~M_{\odot}$. 
The Crab pulsar is described by $M\geq 1.3~M_{\odot}$ and Vela by $M\sim 1.8 M_{\odot}$. 
The XMMU source has a mass of $1.1-1.2~M_{\odot}$. 
All  sources, except the most rapidly cooling objects "C" and "D'' are covered by stars with masses from 1.1 to 1.8 $M_{\odot}$.

In Fig.~\ref{Fig:Cool7a} we show the same as in Fig.~\ref{Fig:Cool3a} and Fig.~\ref{Fig:Cool5a} ($n_c^{\pi}=2~n_0$),   but for the  $1S_0$ $pp$ pairing gaps corresponding to the model BCLL. 
We reproduce the HRC-S Cas~A data for a NS mass $M=1.406~M_\odot$, being smaller than in the cases shown in Fig.~\ref{Fig:Cool3a} ($M=1.695~M_\odot$) and Fig.~\ref{Fig:Cool5a} ($M=1.663~M_\odot$), and having a slightly smaller decline  ($1.6\%$ per decade). 
The whole range of the data is covered by the cooling curves  of  masses from 1.0 to $2.43~M_{\odot}$. 
The Crab pulsar is described by $M\geq 1.1~M_{\odot}$ and Vela by $M\sim 1.55~M_{\odot}$. 
The XMMU source has a mass of $1.0 -1.05 M_{\odot}$ and thus being very light, which could be an argument against this model. 
All  sources, except the most rapidly cooling objects "C" and "D''  are covered by stars with masses from 
1.0 to 1.8 $M_{\odot}$.

In Fig.~\ref{Fig:Cool7b} we show the same as in Fig.~\ref{Fig:Cool3b} and Fig.~\ref{Fig:Cool5b} ($n_c^{\pi}=1.5~n_0$),   but for the  $1S_0$ $pp$ pairing gaps corresponding to the model BCLL. 
We reproduce the HRC-S Cas~A data for a NS mass of $M=1.345~M_\odot$, being smaller than in the cases shown in Fig.~\ref{Fig:Cool3b} ($M=1.660~M_\odot$) and Fig.~\ref{Fig:Cool5b} ($M=1.625~M_\odot$), and obtain a small decline of $1.0\%$ per decade. 
The whole range of the data is covered by cooling curves in a broad interval of  masses, from 0.7 to 
$2.43~M_{\odot}$. 
The Crab pulsar is described by $M\geq 1.0~M_{\odot}$ and Vela by $M\sim 1.5~M_{\odot}$. 
The XMMU source has the mass of $0.7-0.9~M_{\odot}$ being unrealistically light. 
All  sources, except the most rapidly cooling objects "C" and "D''  are covered by stars with masses from 
0.7 to 1.5 $M_{\odot}$.

Thus we conclude that with the DD2 EoS for all three considered choices of the $pp$- gaps  we are able to tune the effective pion gap (for $n_c^{\pi}=1.5-2.5~n_0$) such that all the cooling data are explained rather appropriately. 
If the decline of the Cas~A cooling curve were steeper than $2\%$ it would essentially restrict the different choices. 
In the latter case only choices shown in Figs.~\ref{Fig:Cool3} and~\ref{Fig:Cool5} would prove to be appropriate.  
If the decline of the Cas~A cooling curve is not so steep ($1-2\%$) all three choices of the $pp$ gaps EEHOr, CCYms and BCLL are appropriate, given that  $n_c^{\pi}=1.5-2.5~n_0$ for the EEHOr and CCYms models and $n_c^{\pi}=2-2.5~n_0$ for the BCLL model. 
Other $pp$ gap and effective pion gap choices do not allow to appropriately explain the whole set of the data.

\subsection{NS cooling within DD2vex EoS}

In Fig.~\ref{Fig:Cool8} we show the same as in Fig.~\ref{Fig:Cool7a} ($n_c^{\pi}=2~n_0$, BCLL $pp$ gaps) but for DD2vex EoS. 
We reproduce the ACIS-S Cas~A data for a NS mass $M=1.970~M_\odot$  and obtain a decline of 
$2.6\%$ per decade. The HRC-S  data are described for $M=1.993~M_\odot$.
The whole range of the known cooling data is covered by cooling curves in a broad interval of  masses, from 1.4 to $2.5~M_{\odot}$. 
The Crab pulsar is described by $M\geq 1.6~M_{\odot}$ and Vela by $M\sim 2.1~M_{\odot}$. 
The XMMU source has a mass of $1.4-1.5~M_{\odot}$. 
All  sources, except the most rapidly cooling objects "C" and "D''  are covered by stars with masses from 
1.4 to 2 $M_{\odot}$.

In Fig.~\ref{Fig:Cool8a} we show the same as in Fig.~\ref{Fig:Cool7b} ($n_c^{\pi}=1.5~n_0$, BCLL $pp$ gaps) but for the DD2vex EoS. 
We reproduce the HRC-S Cas~A data for a NS mass $M=1.823~M_\odot$ and obtain a decline of 
$1.3\%$ per 10 years. 
The whole range of the data is covered by cooling curves in a broad interval of  masses, from 1.1 to 
$2.5~M_{\odot}$. 
The Crab pulsar is described by $M\geq 1.4~M_{\odot}$ and Vela by $M\sim 2.0~M_{\odot}$. 
The XMMU source has a mass of $1.1-1.3~M_{\odot}$. 
All  sources, except the most rapidly cooling objects "C" and "D''  are covered by stars with masses from 
1.1 to 1.9 $M_{\odot}$.

In  Fig.~\ref{Fig:Cool8b} we show the case of the AO $pp$ pairing gap and  an effective pion gap corresponding to $n_c^{\pi}=2~n_0$. 
We reproduce the HRC-S Cas~A data for a NS mass of $M=1.740~M_\odot$ and obtain a decline of 
$1.5\%$ per decade. 
The whole range of the data is covered by cooling curves in a broad interval of  masses, from 1.0 to 
$2.5 M_{\odot}$. 
The Crab pulsar is described by $M\geq 1.2 M_{\odot}$ and Vela by $M\sim 2.0~M_{\odot}$. 
The XMMU source has a mass of $1.0-1.15~M_{\odot}$. 
All  sources, except the most rapidly cooling objects "C" and "D'' are covered by stars with masses from 
1.0 to 1.8 $M_{\odot}$.

All other choices of $pp$ gaps and effective pion gaps do not allow to appropriately explain the data. 
The cooling  proves to be too fast in those cases and hot sources cannot be appropriately described 
if the gap drops to zero at smaller densities, as for the  CCYps and BS $pp$ gap choices. 
The cooling is then too slow, and less hot  sources cannot be appropriately described if the gap drops 
for higher densities, as for the EEHO, CCDK, Yak and AV18 gap choices.

The overall description could be improved if other rapid processes, like the pion condensate process, 
or the DU process on hyperons, or the possibility of a quark matter core were included.
This we demonstrate on the example of the AO $pp$ pairing and for the effective pion gap corresponding to $n_c^{\pi}=2.5~n_0$. 
Here, in absence of pion condensation and hyperons the cooling of the rapidly cooling objects can be explained by DU reactions. 
This example is shown in Fig.~\ref{Fig:Cool8c}. 
We reproduce the HRC-S Cas~A data for the NS mass $M=2.230~M_\odot$ and obtain a decline of
$1.1\%$ per decade. 
The whole range of the data is covered by cooling curves in a broad interval of  masses, from 1.0 to 
$2.7~M_{\odot}$. 
The Crab pulsar is described by $M\geq 1.3~M_{\odot}$ and Vela by $M\sim 2.6~M_{\odot}$ 
(with the DU process switched on!). 
The XMMU source has a mass of $1.0-1.2~M_{\odot}$. 
All  sources, except the most rapidly cooling objects "C" and "D''  are covered by stars with masses from 
1.0 to 2.5 $M_{\odot}$. 
The cooling of the cold "C" and "D'' objects is explained by the efficient DU process.

\subsection{CS cooling within DD2vex-QM EoS}

Turning our attention to the scenario of a hybrid star EoS, we have observed in 
Ref.~\cite{Grigorian:2015nva} that the presence of a quark core may lead to an acceleration of the cooling. 
The full set of cooling data can again be described.
In the present work the matching of the  DD2vex hadron EoS with the hNJL quark matter EoS produces the DD2vex-QM  EoS demonstrated in Fig.~\ref{Fig:MR}. 
The quark core appears only for $M>2.08~M_{\odot}$.
As is seen in Fig.~\ref{Fig:Cool8} and Fig.~\ref{Fig:Cool8a} with BCLL $pp$ gaps only the cooling of the "D'' object might be described in terms of the DD2vex-QM model. 
Other objects have lower masses and their cooling is described within the hadronic models.
With the AO $pp$ gaps and for an effective pion gap corresponding to $n_c^{\pi}=2.0~n_0$
the situation is similar and only the "D'' object may be described as a hybrid star with quark matter core. However, for the AO $pp$ gaps and for an effective pion gap corresponding to $n_c^{\pi}=2.5~n_0$, see Fig.~\ref{Fig:Cool8c}, the objects demonstrating an intermediate cooling  
(Cas A, 3C 58, Geminga and RX J1856-3754), as well as the rapidly cooling ones (Vela and CTA-1) 
would have masses above $2.1 M_{\odot}$ and could be considered as hybrid stars in the framework of the DD2vex-QM model.

\section{\label{sec:Remarks}Conclusion}

As is seen in Fig.~\ref{Fig:MR} the neutron star (NS) radii are the larger the stiffer EoS is. 
So, if in the future a very massive compact star (CS), and/or a CS of a middle mass but with a large radius will be observed, this will provide arguments for a stiff EoS, like DD2 or even DD2vex. 
Otherwise, with the presently existing value of the mass for the most massive observed pulsar 
$2.01\pm 0.04 M_{\odot}$ and in absence of the evidence of the CS with radii above 13 km we may 
exploit moderately stiff equations of state (EoS) like the HDD one, yielding $M_{\max}$ only slightly above $2~M_{\odot}$ and radii of $12\pm 1$ km for medium heavy CS. 
Here we should stipulate that we assume that hyperons, delta resonances and pion, kaon and 
charged $\rho-$meson condensates do not appear or, if they appear, soften the EoS only a little.

As we demonstrated,  all three EoS that we used, HDD, DD2 and DD2vex,  are compatible with the existing CS cooling data provided we exploit the nuclear medium cooling scenario developed in our previous works, under the assumption that different sources have different masses. 
The key difference of this scenario compared to the minimal cooling one is that we incorporate the particle-particle hole Fermi liquid effects as well as short-range nucleon repulsion effects. 
Due to that the $NN$ interaction amplitude becomes strongly density dependent being enhanced for 
$n\gsim n_0$ owing to the pion softening effect and suppressed at lower densities owing to correlation effects in particle-particle channel. 
As the result of inclusion of the particle-hole effects, the emissivity of the two-nucleon neutrino processes increases with increasing density in the NS core and the lepton and nucleon heat conductivity terms decrease.

The resulting cooling curves prove to be  sensitive to the value and the density dependence of the $pp$ pairing gap and the effective pion gap. 
In the present work to be specific and to restrict the number of possible choices we disregarded the possibility of the pion condensation assuming a saturation of the pion softening above a critical density, $n_c^{\pi}$, which we varied between 1.5 to  $3.0~n_0$ such that the strongest pion softening occurs for $n_c^{\pi}=1.5~n_0$, cf. Fig.~\ref{Fig:omegatilde}. 
Choosing the $pp$ pairing gap  such that it disappears for sufficiently high densities met in centers  of rather massive stars we are able to reach an overall agreement with the cooling data including Cas~A and the hot source XMMU-J1732, with all three: HDD, DD2 and  DD2vex EoS used in this work. 
We showed that the $pp$ gaps appropriate for a description with the stiffer EoS should drop to zero at lower densities than those yielding  appropriate fits of the cooling data with a softer EoS. 
Allowing for  a stronger decrease of the effective pion gap with increasing density we are able to diminish the value of the mass required for an optimal description of the cooling data for the NS in Cas~A. 
Exploiting the stiffest EoS necessitates the usage of the strongest pion softening.

Fitting a steep decline of the cooling curve for Cas~A compatible with ACIS-S data requires  an appropriate form of the $pp$ pairing gap (model I for the HDD EoS; EEHOr and CCYms for the DD2 EoS and BCLL for the DD2vex EoS, cf. Fig.~\ref{Fig:gaps}) and an efficient medium modified Urca neutrino emissivity obtained with the effective pion gap $n_c^\pi =3.0~n_0$ for the HDD EoS, $n_c^\pi =2.5~n_0$ for the DD2 EoS and $n_c^\pi =2.0~n_0$ for the DD2vex EoS, cf. Fig.~\ref{Fig:omegatilde}.  
The larger the $pp$ gap at the relevant density, the steeper the  decline of the curve describing the  Cas~A cooling data. 
However, with too large $pp$ pairing gaps the masses of the hot objects, like XMMU, and intermediately cooling objects prove to be too high. 
A  rather low value of the lepton contribution to the heat conductivity, being also required to fit the ACIS-S data, follows from the calculations in Ref.~\cite{Shternin:2007ee}, which take electron-electron hole effects into account.

A smaller decline of the cooling curve for Cas~A, like the one being found using data of the HRC-S instrument, together with other cooling data are better reproduced, when one exploits smaller values of the proton gap, like the BCLL one. 
In addition, the above mentioned decline diminishes, if one uses effective pion gaps with a steeper decrease with the density.

To conclude this discussion, exploiting the DD2 EoS, the EEHOr and the CCYms $pp$ pairing gaps and the effective pion gap with $n_c^{\pi}=1.5-2.5~n_0$, as well as using the BCLL proton gaps and  
$n_c^{\pi}=2.0-2.5~n_0$,  allows to appropriately explain the whole set of the cooling data either demonstrating a sharp or a smoother decline of the cooling curve passing through the Cas A data.

With the extremely stiff DD2vex EoS the cooling data are appropriately explained only if one assumes a  strong pion softening  $n_c^{\pi}=1.5~n_0$ and $n_c^{\pi}=2.0~n_0$ with the BCLL model for the $pp$ pairing and $n_c^{\pi}=2.0~n_0$ and $n_c^{\pi}=2.5~n_0$ with the AO model for the $pp$ pairing gaps. 
In both cases the gaps vanish for densities above $2.2 n_0$.

Reference \cite{Klochkov} presented arguments that the mass of the XMMU source is  
$1.2~M_{\odot}<M_{\rm XMMU}<1.8~M_{\odot}$.  
Accepting these limits as a constraint we are able to diminish the number of relevant cooling models. 
The gap of model II for the HDD EoS, the BCLL gap model for the DD2 EoS and the AO gap model for the DD2vex EoS can then be excluded.

Within our scenario the hottest source should be the lightest NS. 
Bearing in mind that most of CS in binary systems have masses $M\simeq 1.35-1.4~M_{\odot}$, we would expect that $M_{\rm XMMU}<1.4~M_{\odot}$. 
Accepting then as a constraint $1.2~M_{\odot}<M_{\rm XMMU}<1.4~M_{\odot}$ we may still reduce the number of appropriate choices retaining the following models:  
{\em for the HDD EoS}: model I of $pp$ gap and the effective pion gap $n_c^{\pi}=3~n_0$, 
Fig.~\ref{Fig:Cool1}; 
{\em for the DD2 EoS}: with the EEHOr gap, $n_c^{\pi}=2~n_0$, Fig.~\ref{Fig:Cool3a}, and 
$n_c^{\pi}=1.5~n_0$, Fig.~\ref{Fig:Cool3b};  
with the CCYms gap, $n_c^{\pi}=2.5~n_0$, Fig.~\ref{Fig:Cool5}, $n_c^{\pi}=2.0~n_0$, 
Fig.~\ref{Fig:Cool5a}, and $n_c^{\pi}=1.5~n_0$, Fig.~\ref{Fig:Cool5b}; and with the BCLL gap, 
$n_c^{\pi}=3.0~n_0$, Fig.~\ref{Fig:Cool6}; and  
{\em for the DD2vex EoS}: with the BCLL gap, $n_c^{\pi}=1.5~n_0$, Fig.~\ref{Fig:Cool8a}.

 Finally, we would like to emphasize that in a self-consistent scheme  the pairing gaps should be computed by taking into account in-medium effects beyond the BCS weak coupling approximation, and for the same interaction that is used to obtain the given  EoS. 
This could reduce number of appropriate models. 
However, the problem cannot be satisfactorily resolved at present.

If we included the possibility of pion condensation for the DD2 EoS, it would appear for 
$M>M_c^{\pi}=2.1~M_{\odot}$ if $n_{c}^{\pi}=3.0~n_0$,  for $M>1.8~M_{\odot}$ if $n_c^{\pi}=2.5~n_0$,  
for $M>1.25 M_{\odot}$ if $n_c^{\pi}=2.0~n_0$, and for $M>0.75~M_{\odot}$ if $n_c^{\pi}=1.5~n_0$, 
see Fig.~\ref{Fig:omegatilde}. 
For the DD2vex EoS, the pion condensation  could appear in our models for $M>2.7~M_{\odot}$ if 
$n_{c}^{\pi}=2.5~n_0$,  for $M>2.1~M_{\odot}$ if $n_c^{\pi}=2.0~n_0$,  and  for $M>1.25~M_{\odot}$ 
if $n_c^{\pi}=1.5~n_0$. 
The cooling curves for the  NS masses below $M_c^{\pi}$ remain the same as in the models without pion condensation, which we have exploited in the present work. 
For $M>M_c^{\pi}$ the cooling becomes more efficient since the pion Urca process switches on. 
This possibility will be studied elsewhere.

There exists the so called ``hyperon puzzle'' \cite{SchaffnerBielich:2008kb,Djapo:2008au} that the filling of the hyperon Fermi seas in CS matter may happen already for $n>2.5-3.0~n_0$ which may result in a significant decrease of the maximum CS mass. 
Moreover, the rapid DU-like processes on the hyperons may occur.
Although recently it was demonstrated \cite{Maslov:2015msa,Maslov:2015wba} how these problems might be smoothened  it would be important to include the hyperons in the cooling code.

\subsection*{Acknowledgments}
We thank C. O. Heinke and E. E. Kolomeitsev for valuable discussions and suggestions. 
D.E. Alvarez-Castillo and S. Typel are gratefully acknowledged for contributing EoS data for the crust and for the DD2vex model.
This research was supported by Narodowe Centrum Nauki (NCN) under grant number
UMO-2014/13/B/ST9/02621, 
by the Munich Institute for Astro- and Particle Physics (MIAPP) of the DFG cluster of excellence 
"Origin and Structure of the Universe"",
and by  the Ministry of Education and Science of the Russian
Federation (Basic part).
H.G. acknowledges support for scientific exchange between JINR Dubna and
Polish Institutes  (by the Bogoliubov - Infeld programme)  and Armenian Institutes (by the Ter-Antonian - Smorodinsky programme). 
The work of D.B. has been supported by the Hessian LOEWE initiative through HIC for FAIR. 
The authors are also grateful for support from the COST Action MP1304 "NewCompStar" for their networking and collaboration activities.

\section*{\label{sec:Appendix}Appendix: Some remarks why  ``minimal cooling" is insufficient.}

Here we formulate reasons why one should go beyond the so-called "minimal cooling" scheme, exploited in most works devoted to the NS cooling (for a review, see \cite{Potekhin:2015qsa}),  where as the most efficient neutrino processes in the NS core are considered the MU processes calculated within the FOPE model \cite{Friman:1978zq} and the neutron $3P_2$ PBF process calculated following Ref.~\cite{Leinson:2009nu}. The NB and $1S_0$ neutron PBF processes are also taken into account.

The FOPE model of the $NN$ interaction suffers from serious drawbacks. 
Indeed, the calculation of the rate of the process can be performed either by exploiting  the squared Born amplitude or the optical theorem, i.e. via the squared matrix element of the diagram or via its imaginary part. 
The imaginary part of the Born amplitude for forward scattering is zero. 
Therefore, the calculation of the emissivity via the imaginary part of the amplitude requires the inclusion of the nucleon-nucleon hole loop diagrams at least up to second order in the coupling $f_{\pi NN}$.  
The FOPE MU result is reproduced if higher order loop terms can be dropped, i.e. only if the one-loop term 
$\propto f_{\pi NN}^2$ is small. 
However, the latter term proves to be numerically not as small already at low densities. 
Indeed, with inclusion 
of this nucleon-nucleon hole contribution into the pion propagator, the pion condensation instability  happens already  for $n=n_c^{\pi}\sim 0.3 n_0$  in isospin symmetric matter \cite{Voskresensky:2001fd}, whereas experiments show that there is no pion condensation in atomic nuclei \cite{Migdal:1990vm}. 
In order to get a correct pion self-energy one should sum up the loop series by taking into account vertices dressed by the loops, which in the Fermi liquid approach depend on the $g_0^{nn}$ and $g_0^{np}$ Landau-Migdal parameters in the spin-spin interaction channel. 
The  repulsive effect of the $NN$ correlations allows to shift $n_c^{\pi}$ above the value of $n_0$ for isospin symmetric matter \cite{Voskresensky:2001fd}. 
Then both calculations, the one via the optical theorem and the other using the Born amplitude (but with MOPE with dressed vertices and dressed pion propagator instead of FOPE) coincide. 
Note that in addition to MOPE there is yet a repulsive short-range term in the $NN$ interaction corrected by the loops which, however, yields a small effect for $n\gsim n_0$, cf. \cite{Voskresensky:2001fd}.

Next, if the pion is treated as free, there should appear a rapid DU-like pionization process, provided the electron chemical potential $\mu_e$ exceeds the pion mass $m_\pi$. 
This process starts at a density much below the nucleon DU threshold. 
Silently, it is disregarded in the minimal cooling schemes, although  this process proves to be forbidden only if one takes into account   in-medium s-wave repulsion, cf. \cite{Migdal:1990vm}.

The inclusion of PBF processes into the minimal cooling scheme, which ignores medium effects in the MU reaction rate, appears to be inconsistent,  since only taking into account the  particle-particle and particle-hole loop diagrams in vertices allows one to preserve the Ward identity and the conservation of the vector current, cf.~\cite{Kolomeitsev:2008mc,Kolomeitsev:2010hr}. 
But then one may ask why one  includes these effects in PBF processes but  ignores similar effects in the two-nucleon processes, like MU?

Now, the electron-electron hole effects are  incorporated in the lepton heat conductivity in the minimal cooling calculations, cf.~\cite{Shternin:2007ee}, resulting in its essential suppression. 
But the same effects are not included in the nucleon term, what again looks inconsequent. 
The PBF process on $1S_0$ paired neutron is included in the minimal cooling scheme by taking into account loop effects in the particle-particle channel with the result that the vector current term is suppressed as $v_{{\rm F}n}^4$, where  $v_{{\rm F}n}$ is the neutron Fermi velocity, 
see~\cite{Leinson:2006gf,Kolomeitsev:2008mc}, and the axial-vector current term is suppressed only as
$v_{{\rm F}n}^2$, see \cite{Kolomeitsev:2008mc,Kolomeitsev:2010hr}. 
But the neutron-neutron hole and electron-electron hole effects are not incorporated into the rate of the  PBF process on the $1S_0$ paired proton what results in an artificial one-two order of magnitude ($c_v^2$) suppression of the rate for this process when compared with that from \cite{Kolomeitsev:2008mc}, where these effects are included.

The loop effects in calculation of the pairing gaps might be very important. 
However, the uncertainty in the results is large because of an exponential dependence of the gap on the poorly known value of the particle-particle interaction in the pairing channel. 
Especially these uncertainties may affect the values of the $3P_2$ gaps, cf.~\cite{Schwenk:2003bc,Khodel:2004nt}. 
Therefore it is difficult to believe that one may  fix the magnitude and the density dependence of the $3P_2$ $nn$ pairing gap in the NS interior from fitting the cooling curves to the existing data, as it was argued by the authors exploiting the minimal cooling scheme, because many at least potentially important effects are ignored  within such an approach. 
The poor knowledge of the density dependence and the amplitudes of the $3P_2$ $nn$ and $1S_0$ $pp$ gaps is a serious barrier for reducing the uncertainties in predictions for all existing cooling schemes.

Concluding, it deems us rather inconsistent to include the loop effects in some processes but to ignore similar effects in other processes. 
These drawbacks are absent in our nuclear medium cooling scheme. 
The price paid for that is a dependence of the quantitative results on the values of the poorly known Fermi liquid parameters in dense NS matter.  
Thus, further experimental and theoretical efforts  are required which will  step by step improve our knowledge of the interaction in dense nuclear matter.


\end{document}